\documentclass[journal=jacsat,manuscript=article]{achemso}

\usepackage{chemformula} % Formula subscripts using \ch{}
\usepackage[T1]{fontenc} % Use modern font encodings
\usepackage{siunitx,physics,float,mathtools}
\usepackage{empheq}
\usepackage{here}
\usepackage{bm}% bold math
\usepackage{xcolor}
\usepackage{natbib}
\usepackage{amsmath,amssymb}
\usepackage{overpic}
\SectionNumbersOn

\author{Yusuke Koide}
\affiliation{ 
Graduate School of Engineering, Nagoya University, Furo-cho, Chikusa, Nagoya, Aichi 464-8603, Japan%\\This line break forced with \textbackslash\textbackslash
}%
\email{koide.yusuke.k1@f.mail.nagoya-u.ac.jp}
\author{Takato Ishida}

\author{Takashi Uneyama}
\author{Yuichi Masubuchi}

\title
  {Steady-state extensional viscosity of wormlike micellar solutions via dissipative particle dynamics simulations}

\abbreviations{IR,NMR,UV}

\begin{document}

% \begin{tocentry}

% Some journals require a graphical entry for the Table of Contents.
% This should be laid out ``print ready'' so that the sizing of the

% Inside the \texttt{tocentry} environment, the font used is Helvetica
% 8\,pt, as required by \emph{Journal of the American Chemical

% The surrounding frame is 9\,cm by 3.5\,cm, which is the maximum
% permitted for  \emph{Journal of the American Chemical Society}
% graphical table of content entries. The box will not resize if the
% content is too big: instead it will overflow the edge of the box.

% This box and the associated title will always be printed on a
% separate page at the end of the document.

% \end{tocentry}

\begin{abstract}

We investigate the steady-state extensional viscosity of wormlike micellar solutions using dissipative particle dynamics simulations. 
As the extension rate increases, the steady-state extensional viscosity initially increases and subsequently decreases after reaching a maximum, as observed in experiments.
We reveal that this nonmonotonic behavior arises from the competition between micellar stretching and scission under uniaxial extensional flow.
We further propose a relation that connects the extensional viscosity to micellar structures and kinetics.
This relation provides a unified description of the extensional viscosity of unentangled wormlike micellar solutions for various temperatures, concentrations, and extension rates.

\end{abstract}

\section{Introduction}

Under certain conditions, surfactants self-assemble into flexible and elongated aggregates known as wormlike micelles.~\cite{Dreiss2007-jf}
Wormlike micelles impart pronounced viscoelasticity to solutions, and this property plays a crucial role in consumer products such as shampoos and body washes.~\cite{Yang2002-ob}
Wormlike micelles also serve as drag-reducing agents in turbulent flows.~\cite{Gyr1995-nn,Zakin1998-zm}
Thus, the rheological properties of wormlike micelles have been an active area of research.
Significant progress has been made in understanding the intriguing behavior of wormlike micellar solutions under shear flow, including shear thickening~\cite{Wunderlich1987-ys,Liu1996-wh} and shear banding.~\cite{Spenley1993-qk,Britton1997-fs}
In addition to shear rheology, extensional rheology is indispensable because wormlike micelles undergo complex flows consisting of shear and extensional modes, for example in turbulent flows. 
Indeed, Lu et al.~\cite{Lu1997-vi} reported that the large extensional viscosity of wormlike micellar solutions was a key property governing the drag reduction performance.

Measuring the extensional rheology of wormlike micellar solutions has been a significant challenge, attracting numerous studies.~\cite{Rothstein2020-uy}
Early studies employed opposed jet devices to generate extensional flows~\cite{Prudhomme1994-tu,Hu1994-dj,Walker1996-cf,Chen1997-xd,Fischer1997-ji,Lin2001-ge,Lu1998-au,Muller2004-hf}.
Walker et al.~\cite{Walker1996-cf} approximated the uniaxial extensional flow field using the Rheometrics RFX device and measured the apparent steady-state extensional viscosity $\eta_{E,\mathrm{app}}(\dot{\epsilon}_\mathrm{app})$ of cetylpyridinium chloride~(CPyCl) and sodium salicylate~(NaSal) solutions as a function of the apparent extension rate $\dot{\epsilon}_\mathrm{app}$.
They identified three regimes of $\eta_{E,\mathrm{app}}(\dot{\epsilon}_\mathrm{app})$: a plateau at low $\dot{\epsilon}_\mathrm{app}$, an increasing trend up to a certain $\dot{\epsilon}_\mathrm{app}$, and a subsequent decrease.
This nonmonotonic behavior of $\eta_{E,\mathrm{app}}(\dot{\epsilon}_\mathrm{app})$ has been observed in various surfactant solutions~\cite{Prudhomme1994-tu,Hu1994-dj,Chen1997-xd,Fischer1997-ji,Lin2001-ge}.
Chen and Warr~\cite{Chen1997-xd} conducted the light scattering measurement in extensional flows to evaluate the gyration radius $R_{g}$ of micelles.
They demonstrated that $R_{g}$ exhibited a nonmonotonic behavior similar to $\eta_{E,\mathrm{app}}(\dot{\epsilon}_\mathrm{app})$, implying that the decrease in $\eta_{E,\mathrm{app}}(\dot{\epsilon}_\mathrm{app})$ can be partly attributed to micellar scission.
However, $\eta_{E,\mathrm{app}}(\dot{\epsilon}_\mathrm{app})$ obtained with opposed jet devices can be subject to non-ideal effects, such as the presence of shear flows and liquid inertia~\cite{Dontula1997-oa}.
Indeed, Haward et al.~\cite{Haward2012-pq} examined the velocity field within a microfluidic cross-slot device at high extension rates and argued that the reduction of $\eta_{E,\mathrm{app}}(\dot{\epsilon}_\mathrm{app})$ obtained with opposed jets was mainly due to a flow instability.
Filament-stretching rheometers, which achieve a nearly ideal homogeneous uniaxial extension~\cite{McKinley2002-ck}, were then used to study the extensional rheology of wormlike micellar solutions~\cite{Rothstein2003-op,Bhardwaj2007-yd,Bhardwaj2007-zk,Chellamuthu2008-sm}.
Rothstein~\cite{Rothstein2003-op} investigated the transient extensional properties of cetyltrimethylammonium bromide~(CTAB) and NaSal solutions and reported that wormlike micellar solutions exhibited a significant strain hardening and experienced a filament rupture at a critical stress.
They conjectured that this macroscopic rupture arose from micellar scission occurring at microscopic scales.
In addition, the maximum extensional viscosity attained before filament rupture exhibited a similar behavior to $\eta_{E,\mathrm{app}}(\dot{\epsilon}_\mathrm{app})$ measured at large $\dot{\epsilon}_\mathrm{app}$ in opposed jet experiments~\cite{Prudhomme1994-tu,Walker1996-cf,Chen1997-xd}.
Recent studies have attempted to measure the extensional rheology of dilute wormlike micellar solutions with low shear viscosities using several techniques, including the capillary break-up extensional rheometer~(CaBER)~\cite{Yesilata2006-rx,Bhardwaj2007-zk,Chellamuthu2008-sm,Miller2009-jo,Sachsenheimer2014-uz,Omidvar2018-es,Wu2018-np,Omidvar2019-rz}, the dripping-onto-substrate~(DoS)~\cite{Wu2018-np,Omidvar2019-rz,Fukushima2022-zi}, and the liquid dripping method~\cite{Tamano2017-og}.
However, Gaillard et al.~\cite{Gaillard2024-ds} demonstrated that the extensional relaxation time obtained with CaBER depended on the plate diameter and the droplet volume. 
Wu and Mohammadigoushki~\cite{Wu2020-vd} reported that in DoS experiments, extensional properties can be affected by the dynamics of the contact line.
Thus, accurately measuring the extensional rheology of wormlike micellar solutions, especially with low viscosities, remains challenging.

Molecular simulations can complement experimental measurements of the extensional rheology of wormlike micelles because they can realize ideal extensional flows and provide molecular-level insights.
Several studies have simulated micelles under uniaxial extensional flow.
Dhakal and Sureshkumar~\cite{Dhakal2016-yk} conducted coarse-grained molecular dynamics~(CGMD) simulations of a rodlike micelle and a U-shaped micelle under uniaxial extensional flow and reported that micellar scission occurred above a critical extension rate.
Mandal and Larson~\cite{Mandal2018-ki} investigated the scission properties of CTAB/NaSal micelles using the CGMD method.
They demonstrated that the breaking stress of micelles was consistent with previous experimental results.~\cite{Rothstein2003-op}
However, the conventional method for applying extensional flows, where the simulation box is simply deformed according to a given flow field, fails at a finite strain due to the collapse of the simulation box.
This limitation hinders the comprehensive investigation of the steady-state properties of wormlike micelles under uniaxial extensional flow.
Thus, considerable uncertainty still exists regarding the steady-state extensional viscosity $\eta_{E}(\dot{\epsilon})$ of wormlike micellar solutions, including its behavior at high extension rates, which are challenging to measure experimentally, and the molecular-level mechanism underlying the nonmonotonic dependence of $\eta_{E}(\dot{\epsilon})$.

In the case of planar extensional flows, the difficulty of simulating extensional flows at arbitrarily large strains has been overcome by the Kraynik--Reinelt~(KR) boundary condition~\cite{Kraynik1992-fa}.
Recently, the generalized KR~(GKR) boundary condition, which is a three-dimensional extension of the KR boundary condition, was developed~\cite{Dobson2014-kr,Hunt2016-vl}.
This method allows long-time simulations of extensional flows by systematically remapping the simulation box.
The advent of this technique has led to molecular simulations of complex fluids, such as polymer melts, under extensional flow, and these simulations have revealed their intriguing flow-induced behavior~\cite{Nicholson2016-aj,O-Connor2018-mh,OConnor2019-po,Murashima2021-wh,Murashima2022-kd}.

In the present study, we aim to elucidate the steady-state extensional properties of wormlike micelles by applying the GKR method to surfactant solutions.
For this purpose, we employ the dissipative particle dynamics~(DPD) method~\cite{Hoogerbrugge1992-ng,Espanol1995-mx}.
Due to coarse-graining and soft-core potentials, the DPD method enables long-timescale simulations of wormlike micelles at relatively low computational cost~\cite{Yamamoto2005-sl,Arai2007-gs,Wang2018-db,Wand2020-qm,Koide2022-bp,Koide2023-ao,Hendrikse2023-im,Koide2025-sw}.
Previous studies~\cite{Koide2022-bp,Koide2023-ao} conducted DPD simulations of surfactant solutions under uniform shear flow and investigated the shear viscosity as well as micellar scission and alignment.
Here, we impose a steady uniaxial extensional flow on the same system and examine the steady-state extensional viscosity $\eta_{E}(\dot{\epsilon})$ along with the micellar structure and dynamics.
We demonstrate that $\eta_{E}(\dot{\epsilon})$ exhibits a nonmonotonic dependence on $\dot{\epsilon}$, as observed in experiments~\cite{Prudhomme1994-tu,Hu1994-dj,Walker1996-cf,Chen1997-xd,Fischer1997-ji,Lin2001-ge}, and this nonmonotonicity arises from the competition between micellar stretching and scission.

\section{Method}
\subsection{Dissipative particle dynamics simulations of surfactant solutions}

\begin{figure}
  \centering
  \begin{overpic}[width=0.5\linewidth]{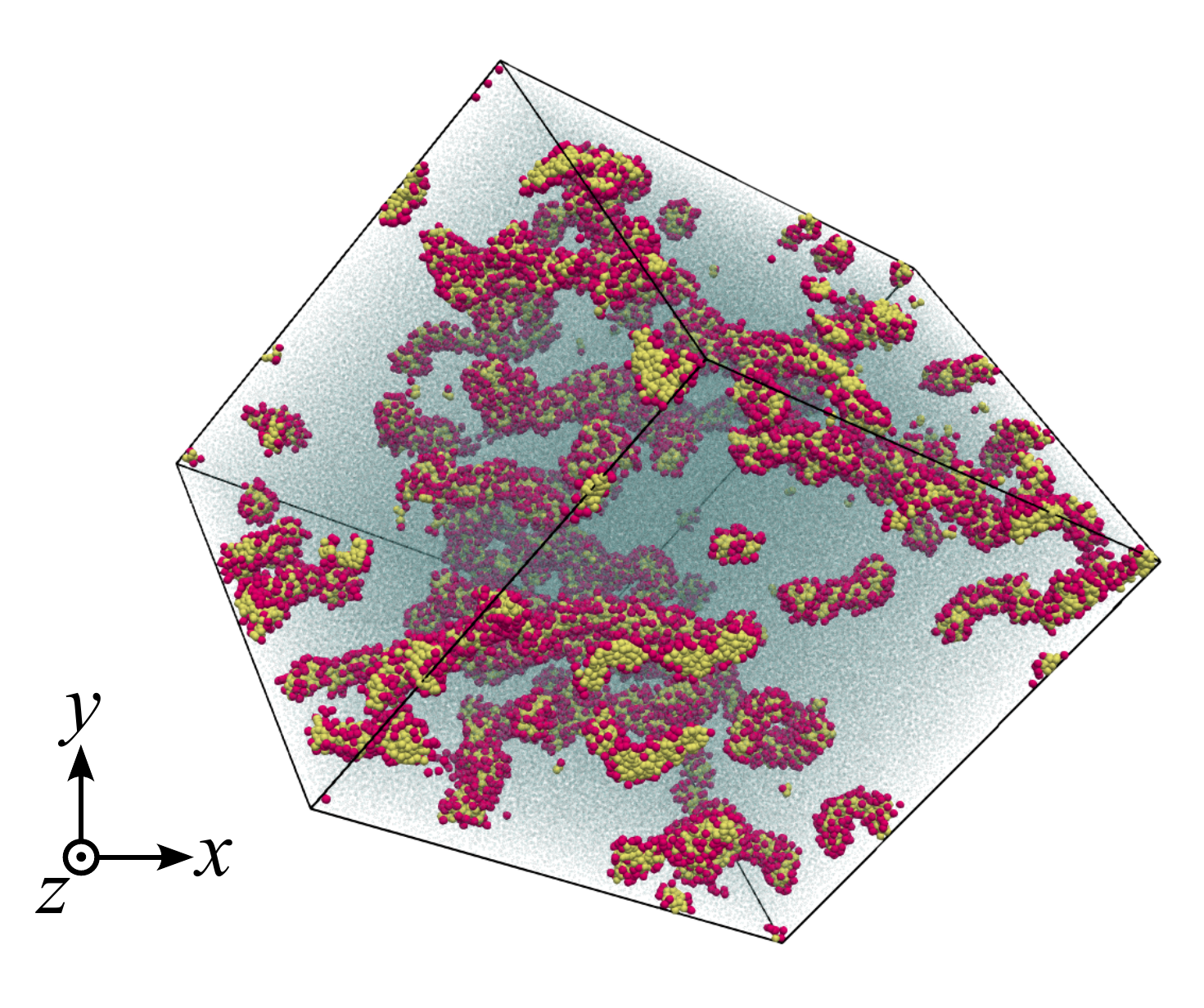} 
  \end{overpic}
  \caption{Snapshot of the surfactant solution for $k_BT=1$ and $\phi=0.05$ under uniaxial extensional flow with $\dot{\epsilon}=0.002$. Hydrophilic and hydrophobic particles are indicated in red and yellow, respectively. For clarity, water particles are represented by blue dots.}
  \label{fig:snapshot}
\end{figure}

We conduct DPD simulations of nonionic surfactant solutions under uniaxial extensional flow, as shown in Fig.~\ref{fig:snapshot}.
In the DPD method, each particle represents a group of atoms or molecules and obeys Newton's laws of motion.
The interactions between DPD particles involve four types of forces.
Repulsive, dissipative, and random forces act between any two particles.
For surfactant molecules, which consist of one hydrophilic head particle and two hydrophobic tail particles, bonded forces are added to connect them.
We use the harmonic bond force $\bm{F}_{ij}^\mathrm{B}$ expressed as
\begin{equation}
    \bm{F}_{ij}^\mathrm{B} = -k_s (|\bm{r}_{ij}|-r_\mathrm{eq}) \bm{e}_{ij}, \label{eq:bond_force} 
\end{equation}
where $k_s$ is the spring constant, $r_\mathrm{eq}$ is the equilibrium bond distance, $\bm{r}_{ij}=\bm{r}_i-\bm{r}_j$, and $\bm{e}_{ij}=\bm{r}_{ij}/|\bm{r}_{ij}|$ with $\bm{r}_i$ being the position of the $i$-th particle.
Further details of the simulation protocol can be found in previous publications~\cite{Koide2022-bp,Koide2023-ao}.
In the following, all quantities are nondimensionalized by $k_BT_0$, $m$, and $r_c$, where $k_B$ is the Boltzmann constant, $T_0$ is the reference temperature, $m$ is the mass of the DPD particle, and $r_c$ is the cutoff distance.

%   --------------------
%   --------------------

In the present study, the parameters of DPD simulations are set as follows:
the total number of particles is $N=648\,000$; the number density of particles is $\rho=3$; the random force coefficient is $\sigma=3$; the spring constant is $k_s=50$; the equilibrium bond distance is $r_\mathrm{eq}=0.8$; the repulsive force coefficients between different types of particles are $a_\mathrm{hh}=25$, $a_\mathrm{ht}=60$, $a_\mathrm{hw}=20$, $a_\mathrm{tt}=25$, $a_\mathrm{tw}=60$, and $a_\mathrm{ww}=25$, where h, t, and w denote head, tail, and water particles, respectively.
These values of $a_\mathrm{ij}$ follow a previous study~\cite{Li2019-wy}, where surfactants form wormlike micelles above a certain volume fraction $\phi$.
Due to the soft-core potential employed in the DPD method, our simulations do not capture entanglement effects, as previously noted for polymer systems.~\cite{Pan2002-oq}
We explore the extensional rheology of unentangled wormlike micellar solutions by changing $\phi$ and $T$.
When $T$ is changed, the dissipative force coefficient $\gamma$ is adjusted to satisfy the fluctuation-dissipation relation $\sigma^2=2\gamma k_BT$.
We have confirmed that varying $\sigma$ instead of $\gamma$ yields nearly identical results within the considered range of $T$.

\subsection{Uniaxial extensional flow}
\label{subsec:extensional_flow}

We use the SLLOD equations\cite{evans_morriss_2008} and the GKR boundary condition~\cite{Dobson2014-kr,Hunt2016-vl} to impose homogeneous uniaxial extensional flow at arbitrarily large strains.
This approach has been theoretically justified by Daivis and Todd~\cite{Daivis2006-pq,Todd2007-ps}.
The SLLOD equations read 
\begin{align}
  &\frac{d{\bm{r}_i}}{dt} = {\bm{p}_i} + (\nabla \bm{u})^\mathsf{T} \cdot\bm{r}_i \\
  &\frac{d{\bm{p}}_i}{dt} = \bm{F}_i - (\nabla \bm{u})^\mathsf{T} \cdot\bm{p}_i ,
\label{eq:SLLOD_equation}
\end{align}
where $\bm{p}_i$ is the so-called peculiar momentum of the $i$-th particle, $\bm{F}_i$ is the force acting on that particle, and $(\nabla\bm{u})_{ij}=\partial u_j/\partial x_i$ is the velocity gradient tensor with $\bm{u}$ representing the velocity of the imposed flow field.
We calculate the dissipative force, which depends on the relative velocity of interacting particles, by using $\bm{p}_i$ instead of the laboratory momentum.
For the steady uniaxial extensional flow, $\nabla\bm{u}$ is expressed as 
\begin{equation}
  \nabla\bm{u} = \begin{pmatrix}
      \dot{\epsilon}&0&0\\
      0&-\dot{\epsilon}/2&0\\
      0&0&-\dot{\epsilon}/2
  \end{pmatrix},
  \label{eq:uniaxial}
\end{equation}
where $\dot{\epsilon}$ is the extension rate.
Under this extensional flow, the $i$-th lattice basis vector $\bm{e}_i$ of the simulation box deforms as 
\begin{equation}
  \frac{d}{dt}\bm{e}_i = (\nabla \bm{u})^\mathsf{T} \cdot \bm{e}_i\label{eq:box_deform}.
\end{equation}
To prevent the collapse of the simulation box at large strains, we apply the GKR method~\cite{Dobson2014-kr,Hunt2016-vl}, which systematically remaps the deformed simulation box.
We also use Semaev's algorithm~\cite{Semaev2001-sn} to remap the simulation box more efficiently~(i.e., for finding a less deformed simulation box).
To prevent instabilities caused by the finite numerical precision~\cite{Todd2000-rh}, the total peculiar momentum is periodically reset to zero, as in previous studies~\cite{Hunt2016-vl,Nicholson2016-aj}.
We employ the dynamic size cell list algorithm to reduce the computational cost of the short-range interactions under uniaxial extensional flow.~\cite{Dobson2016-zk}

The time integration is performed with the modified velocity Verlet method.~\cite{Groot1997-je}
Here, the parameter $\lambda$ introduced in this scheme and the time step $\Delta t$ are set to $\lambda=0.65$ and $\Delta t=0.04$, respectively.
These parameters result in sufficiently accurate temperature control, although the temperature rises slightly under extensional flows with large $\dot{\epsilon}$~(see Appendix A).
The initial simulation box is a cube with dimensions $60\times 60\times 60$.
We conduct equilibrium simulations for $20,000$ time units from a random initial configuration until the potential energy and the number of micelles reach statistically steady values.
After this initial equilibration, we impose uniaxial extensional flows using the GKR method and SLLOD equations.
Since the present study focuses on the steady-state extensional viscosity, all analyses are performed $4\,000$ time units after the onset of uniaxial extensional flows.
Given that the longest relaxation time for all cases is below $1\,500$ time units, both the initial equilibration and the waiting time after imposing flows are sufficiently long.
All DPD simulations are conducted with our in-house code.

\section{Results}
\subsection{Steady-state extensional viscosity}
\begin{figure*}
  \centering
      \begin{tabular}{c}
      \begin{minipage}{0.5\hsize}
          \begin{overpic}[width=1\linewidth]{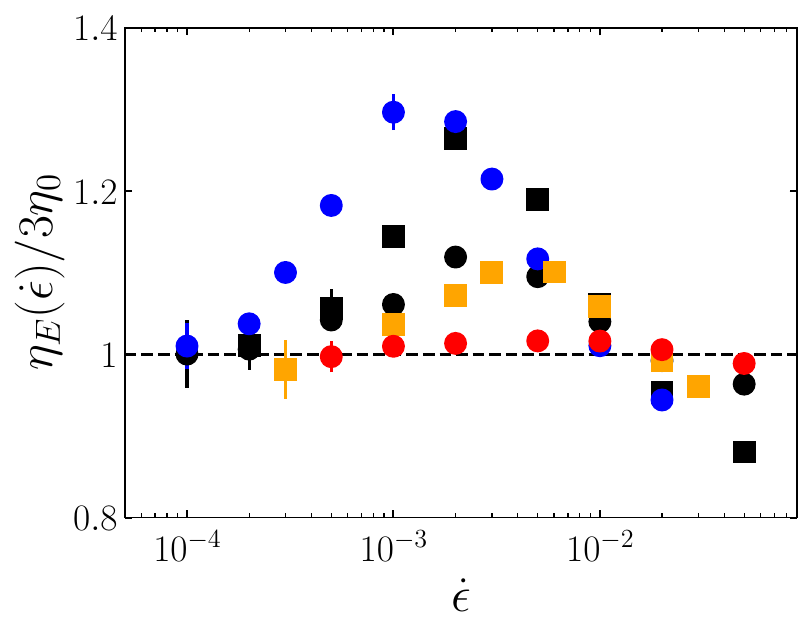}
              \linethickness{3pt}
        \put(0,70){(a)}

          \end{overpic}
      \end{minipage}
      \begin{minipage}{0.5\hsize}
          \begin{overpic}[width=1\linewidth]{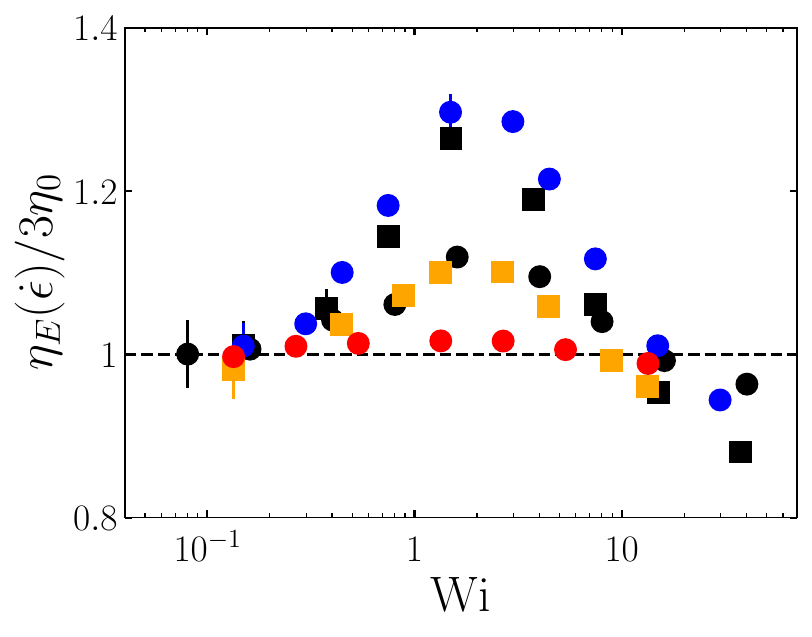}
              \linethickness{3pt}
        \put(0,70){(b)}

          \end{overpic}
      \end{minipage}
      \end{tabular}
  \caption{Steady-state extensional viscosity $\eta_E(\dot{\epsilon})$ normalized by three times the zero-shear viscosity $3\eta_0$ as a function of (a) the extension rate $\dot{\epsilon}$ and (b) the Weissenberg number $\mathrm{Wi}$. Different colors denote different values of $k_BT$: blue, $k_BT=0.9$; black, $1$; orange, $1.1$; red, $1.2$. Different symbols denote different values of the surfactant volume fraction $\phi$: circle, $\phi=0.05$; square, $0.1$. The dashed lines indicate the Newtonian limit $\eta_E(\dot{\epsilon})/3\eta_0=1$.}

      \label{fig:vis}
\end{figure*}

In this subsection, we investigate the extension-rate dependence of the steady-state extensional viscosity of micellar solutions for various concentrations and temperatures.
We define the steady-state extensional viscosity $\eta_E(\dot{\epsilon})$ as 
\begin{equation}
  \eta_E(\dot{\epsilon}) = \frac{\sigma_{xx}-(\sigma_{yy}+\sigma_{zz})/2}{\dot{\epsilon}}, \label{eq:viscosity}
\end{equation}
where $\sigma_{\alpha\beta}$ is the stress tensor.
Here, we employ the Irving--Kirkwood stress tensor~\cite{Irving1950-lc,Liu2015-yj} to calculate $\sigma_{\alpha\beta}$:
\begin{equation}
  \sigma_{\alpha\beta} = -\frac{1}{V}\left\langle \sum_{i<j} r_{ij,\alpha}F_{ij,\beta}+\sum_{i} p_{i\alpha}p_{i\beta}\right \rangle, 
\end{equation}
where $V$ is the volume of the system and $\langle\cdot \rangle$ denotes the ensemble average in the steady state.
Figure~\ref{fig:vis} shows $\eta_E(\dot{\epsilon})$ normalized by three times the zero shear viscosity $3\eta_0$ as a function of $\dot{\epsilon}$ for various values of the surfactant volume fraction $\phi$ and $k_BT$.
Note that we evaluate $\eta_0$ by conducting additional DPD simulations of the same systems under shear flow with very low shear rates.~\cite{Koide2022-bp}
We confirm that $\eta_E(\dot{\epsilon})/3\eta_0$ approaches $1$ as $\dot{\epsilon}\to 0$ regardless of $\phi$ and $k_BT$, which is consistent with the expected Newtonian limit. 
As $\dot{\epsilon}$ increases, $\eta_E(\dot{\epsilon})/3\eta_0$ first increases and subsequently decreases after reaching the maximum value.
This nonmonotonic behavior of $\eta_E(\dot{\epsilon})/3\eta_0$ is consistent with previous observations in experiments\cite{Prudhomme1994-tu,Hu1994-dj,Walker1996-cf,Chen1997-xd,Fischer1997-ji,Lin2001-ge}.
Incidentally, the shear viscosity $\eta(\dot{\gamma})$ of the considered surfactant solutions exhibits only shear-thinning behavior.~\cite{Koide2022-bp}

To compare $\dot{\epsilon}$ dependence of $\eta_E(\dot{\epsilon})$ among different systems, we introduce the Weissenberg number $\mathrm{Wi}=\tau_{\Lambda}\dot{\epsilon}$ defined as the product of the longest relaxation time $\tau_{\Lambda}$ of micelles and the extension rate $\dot{\epsilon}$.
We evaluate $\tau_{\Lambda}$ from the aggregation number $N_\mathrm{ag}$ dependence of the rotational relaxation time $\tau_r(N_\mathrm{ag})$ and the average lifetime $\tau_b(N_\mathrm{ag})$ of micelles in equilibrium using the method proposed in a previous study~\cite{Koide2022-bp}.
We provide a detailed explanation in Sec.~\ref{sec:discussion}.
Figure~\ref{fig:vis}(b) shows $\eta_E(\dot{\epsilon})/3\eta_0$ as a function of $\mathrm{Wi}$.
We observe that $\mathrm{Wi}$ dependence of $\eta_E(\dot{\epsilon})/3\eta_0$ qualitatively exhibits similar behavior, regardless of $\phi$ and $k_BT$.
Specifically, $\eta_E(\dot{\epsilon})/3\eta_0$ begins to increase above $\mathrm{Wi}\simeq 0.3$ and reaches its maximum at $\mathrm{Wi}\simeq 2$.
A further increase in $\mathrm{Wi}$ leads to a decrease in $\eta_E(\dot{\epsilon})/3\eta_0$.
Walker et al.~\cite{Walker1996-cf}  used an opposed jet device and reported the apparent extensional viscosity $\eta_{E,\mathrm{app}}(\dot{\epsilon}_\mathrm{app})$ of CPyCl/NaSal solutions as a function of the product $\dot{\epsilon}_\mathrm{app}\tau_R$ of the apparent extension rate $\dot{\epsilon}_\mathrm{app}$ and the rheological relaxation time $\tau_R$. 
They observed that $\eta_{E,\mathrm{app}}(\dot{\epsilon}_\mathrm{app})$ began to increase at $\dot{\epsilon}_\mathrm{app}\tau_R\simeq 0.7$ and reached its maximum at $\dot{\epsilon}_\mathrm{app}\tau_R=4\text{--}7$ depending on the temperature and the concentration.
Because the value of $\mathrm{Wi}$ at which $\eta_E(\dot{\epsilon})/3\eta_0$ attains its maximum is closely related to flow-induced scission of micelles~(see the next subsection), the location of the maximum is nonuniversal and may depend on micellar properties, especially scission kinetics.

We also observe that $\mathrm{Wi}$ dependence of $\eta_E(\dot{\epsilon})/3\eta_0$ quantitatively varies with $\phi$ and $k_BT$.
For $k_BT=1$, $\eta_E(\dot{\epsilon})/3\eta_0$ at $\phi=0.1$~(black squares) exhibits a more significant change than that at $\phi=0.05$~(black circles).
Since surfactants form larger micelles at a larger value of $\phi$~[see Fig.~\ref{fig:pdf_nag}(b)], $\eta_E(\dot{\epsilon})/3\eta_0$ at $\phi=0.1$ is more susceptible to the effect of extensional flows.
Regarding $k_BT$ dependence, $\mathrm{Wi}$ dependence of $\eta_E(\dot{\epsilon})/3\eta_0$ becomes more moderate as $k_BT$ increases.
In particular, $\eta_E(\dot{\epsilon})/3\eta_0$ at $k_BT=1.2$~(red circles) hardly changes as $\mathrm{Wi}$ increases.
In the following, we explain the observed $\phi$ and $k_BT$ dependence of $\eta_E(\dot{\epsilon})/3\eta_0$ in an integrated way from the perspective of micellar structures and kinetics~(see Sec.~\ref{sec:discussion}).
% %   --------------------
%   --------------------

\subsection{Flow-induced scission of micelles}
\label{subsec:scission}

\begin{figure*}
  \centering
      \begin{tabular}{c}
      \begin{minipage}{0.5\hsize}
          \begin{overpic}[width=1\linewidth]{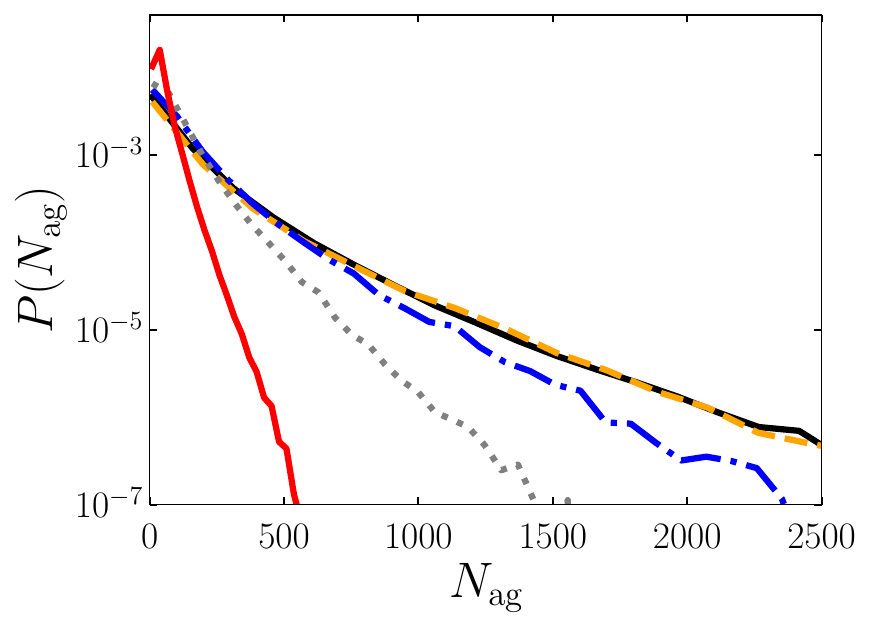}
              \linethickness{3pt}
                \put(0,65){(a)}

          \end{overpic}
      \end{minipage}
      \begin{minipage}{0.5\hsize}
          \begin{overpic}[width=1\linewidth]{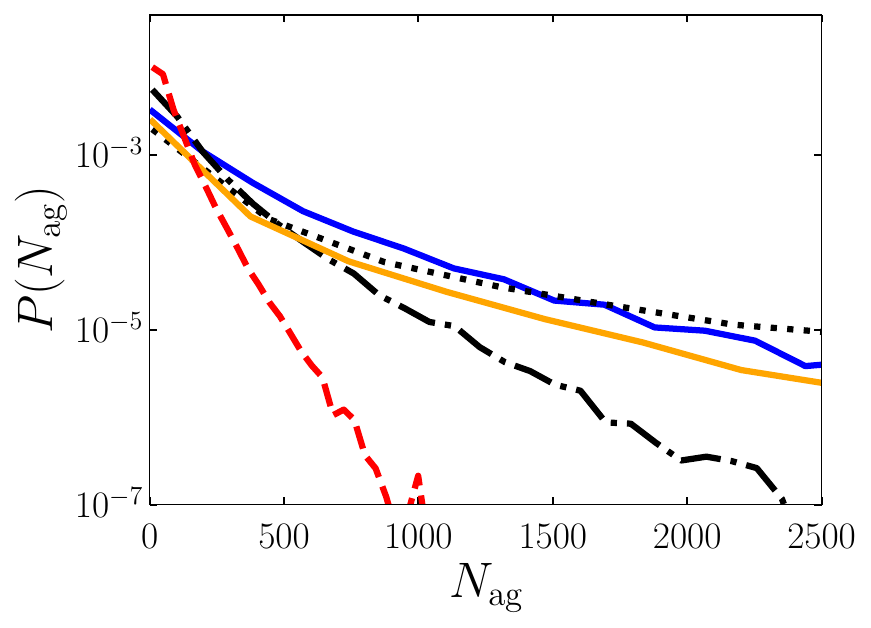}
              \linethickness{3pt}
                \put(0,65){(b)}

          \end{overpic}
      \end{minipage}
      \end{tabular}
      \caption{Probability density function $P(N_\mathrm{ag})$ of the aggregation number $N_\mathrm{ag}$. In (a), different lines denote different values of $\mathrm{Wi}$ for $k_BT=1$ and $\phi=0.05$: black solid line, $\mathrm{Wi}=0$; orange dashed line, $0.16$; blue dash-dotted line, $1.6$; gray dotted line, $4$; red solid line, $16$.
       In (b), different lines denote different values of $\phi$ and $k_BT$ for $\mathrm{Wi}\simeq 2$: blue solid line, $(k_BT,\phi)=(0.9,0.05)$; black dash-dotted line, $(1,0.05)$; black dotted line, $(1,0.1)$; orange solid line, $(1.1,0.1)$; red dashed line, $(1.2,0.05)$.}

      \label{fig:pdf_nag}
\end{figure*}

This subsection examines flow-induced scission of micelles, which is relevant to the decrease in the extensional viscosity.
For this purpose, we define a micelle using a method employed in previous studies.~\cite{Vishnyakov2013-ge,Lee2016-cx}
In this method, two surfactant molecules belong to the same cluster if a hydrophobic particle of one surfactant molecule is within $r_c(=1)$ of a hydrophobic particle of the other.
If a cluster has an aggregation number $N_\mathrm{ag}$ larger than a threshold value $n_\mathrm{mic}(=10)$, the cluster is regarded as a micelle.
We have verified that within the range $5\leq n_\mathrm{mic} \leq 20$, variations in $n_\mathrm{mic}$ have negligible effects on the following results.
Figure~\ref{fig:pdf_nag}(a) shows the probability density function~(PDF) $P(N_\mathrm{ag})$ of the aggregation number $N_\mathrm{ag}$ of micelles for various $\mathrm{Wi}$ with $k_BT=1$ and $\phi=0.05$.
For $\mathrm{Wi}=0.16$, $P(N_\mathrm{ag})$ almost coincides with that at equilibrium, indicating that flow-induced scission hardly occurs.
In contrast, for $\mathrm{Wi}=1.6$, $P(N_\mathrm{ag})$ at large $N_\mathrm{ag}$ decreases, and this tendency becomes more evident as $\mathrm{Wi}$ increases.
This decrease in the population of large micelles results from flow-induced scission of micelles.
Figure~\ref{fig:pdf_nag}(b) shows $P(N_\mathrm{ag})$ at $\mathrm{Wi}\simeq 2$, where $\eta_E(\dot{\epsilon})$ reaches its maximum value, for various values of $\phi$ and $k_BT$.
We observe that $P(N_\mathrm{ag})$ for $k_BT=1.2$ exhibits a smaller value than $P(N_\mathrm{ag})$ for $k_BT=0.9$ at large $N_\mathrm{ag}$ .
Since $\phi$ is common in these two systems, relatively small micelles are prevalent in the system with $k_BT=1.2$.
Qualitatively, this difference in $P(N_\mathrm{ag})$ contributes to $k_BT$ dependence of the maximum value of $\eta_E(\dot{\epsilon})/3\eta_0$~(Fig.~\ref{fig:vis}).
However, a notable exception is observed between the systems with $(k_BT,\phi)=(1,0.05)$ and $(1.1,0.1)$.
Despite the latter system having larger values of $P(N_\mathrm{ag})$ for $N_\mathrm{ag}\gtrsim 500$, $\eta_E(\dot{\epsilon})/3\eta_0$ exhibits similar behavior in the two systems.
As will be discussed in Sec.~\ref{sec:discussion}, the aggregation-number distribution alone is insufficient to explain the quantitative behavior of $\eta_E(\dot{\epsilon})$.
% %   --------------------
%   --------------------

\begin{figure}
  \centering
  \begin{overpic}[width=0.5\linewidth]{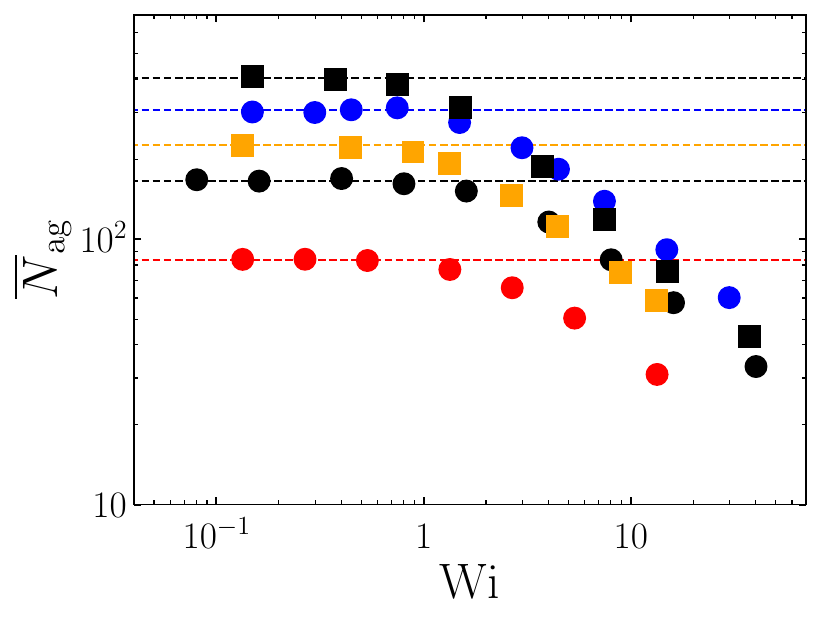} 
  \end{overpic}
  \caption{Mean aggregation number $\overline{N}_\mathrm{ag}$ as a function of the Weissenberg number $\mathrm{Wi}$. The colors and symbols are the same as in Fig.~\ref{fig:vis}. The dashed lines indicate the values of $\overline{N}_\mathrm{ag}$ at equilibrium, with the colors corresponding to the values of $k_BT$. The error bars denote the standard deviations from three independent simulations and are smaller than the symbol size.}
  \label{fig:mean_nag}
\end{figure}

To systematically compare $\mathrm{Wi}$ dependence of $\eta_E(\dot{\epsilon})$ and flow-induced scission, we evaluate the mean aggregation number $\overline{N}_\mathrm{ag}$ of micelles.
Figure~\ref{fig:mean_nag} shows $\overline{N}_\mathrm{ag}$ as a function of $\mathrm{Wi}$ for different values of $k_BT$ and $\phi$.
We observe that $\overline{N}_\mathrm{ag}$ hardly differs from the value at equilibrium for $\mathrm{Wi}\lesssim 2$, whereas $\overline{N}_\mathrm{ag}$ decreases with increasing $\mathrm{Wi}$ for $\mathrm{Wi}\gtrsim 2$.
This demonstrates that extensional flows promote micellar scission above $\mathrm{Wi}\simeq 2$ regardless of the values of $k_BT$ and $\phi$.
Since $\eta_E(\dot{\epsilon})$ also begins to decrease at $\mathrm{Wi}\simeq 2$~[Fig.~\ref{fig:vis}(b)], flow-induced scission of micelles causes the decreasing tendency of $\eta_E(\dot{\epsilon})$ for $\mathrm{Wi}\gtrsim 2$.
%   --------------------
%   --------------------

\begin{figure*}
  \centering
      \begin{tabular}{c}
      \begin{minipage}{0.5\hsize}
          \begin{overpic}[width=1\linewidth]{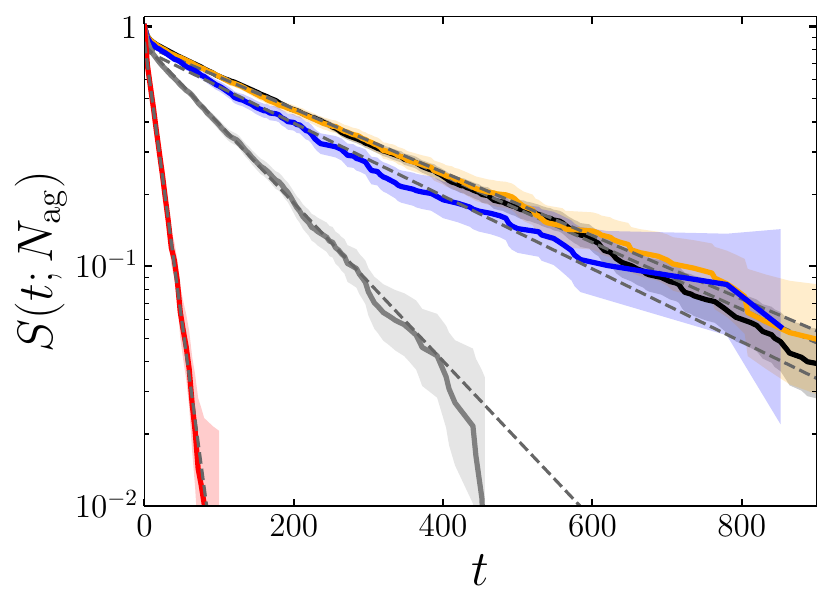}
              \linethickness{3pt}
        \put(0,70){(a)}

          \end{overpic}
      \end{minipage}
      \begin{minipage}{0.5\hsize}
          \begin{overpic}[width=1\linewidth]{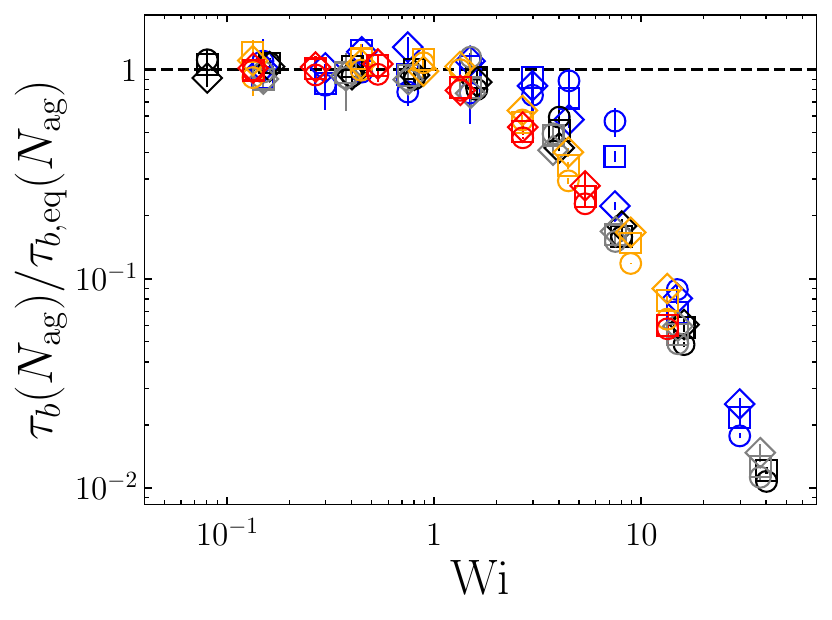}
              \linethickness{3pt}
        \put(0,70){(b)}

          \end{overpic}
      \end{minipage}
      \end{tabular}
      \caption{(a) Survival function $S(t;N_\mathrm{ag})$ of micelles with $N_\mathrm{ag}=300$ with the $95\%$ confidence interval for $\mathrm{Wi}=0$~(black), $0.16$~(orange), $1.6$~(blue), $4$~(gray), and $16$~(red). The gray dashed lines indicate an exponential fit to $S(t;N_\mathrm{ag})$. (b) Average lifetime $\tau_b(N_\mathrm{ag})$ of micelles with $N_\mathrm{ag}=150$~(circle), $200$~(square), and $300$~(diamond) normalized by the value $\tau_{b,\mathrm{eq}}(N_\mathrm{ag})$ at equilibrium as a function of the Weissenberg number $\mathrm{Wi}$ for $(k_BT,\phi)=(0.9,0.05)$~(blue), $(1,0.05)$~(black), $(1,0.1)$~(gray), $(1.1,0.1)$~(orange), and $(1.2,0.05)$~(red). The black dashed line indicates $\tau_{b}(N_\mathrm{ag})/\tau_{b,\mathrm{eq}}(N_\mathrm{ag})=1$. The error bars denote the standard deviations from three independent simulations.}

      \label{fig:lifetime}
\end{figure*}

We explore the detailed scission properties of individual micelles by focusing on their lifetime, defined as the time interval between micellar birth and scission.
Following the same protocol adopted in previous studies~\cite{Koide2022-bp,Koide2023-ao}, scission is defined as an event in which $N_\mathrm{ag}$ of a micelle decreases by more than the threshold value $n_\mathrm{mic}(=10)$ within a given time interval $\delta t(=100\Delta t)$.
To evaluate the statistical properties of micellar scission for each $N_\mathrm{ag}$, we estimate the survival function $S(t;N_\mathrm{ag})$ from lifetimes of micelles whose aggregation numbers are within the range $[N_\mathrm{ag}-\Delta N_\mathrm{ag}/2,N_\mathrm{ag}+\Delta N_\mathrm{ag}/2]$, using the Kaplan--Meier method~\cite{Kaplan1958-yn}.
In the present study, we set $\Delta N_\mathrm{ag}$ to $0.1N_\mathrm{ag}$.
It is worth emphasizing that as mentioned in Sec.~\ref{subsec:extensional_flow}, lifetime data are collected in statistically steady states under uniaxial extensional flow.
Figure~\ref{fig:lifetime}(a) shows $S(t;N_\mathrm{ag})$ of micelles with $N_\mathrm{ag}=300$ for various $\mathrm{Wi}$.
As reported in a previous study~\cite{Koide2022-bp}, $S(t;N_\mathrm{ag})$ exhibits an exponential decay for $\mathrm{Wi}=0$~(i.e., in equilibrium), indicating that scission of individual micelles occurs with a fixed probability per unit time.
We observe that $S(t;N_\mathrm{ag})$ remains unchanged for $\mathrm{Wi}=0.16$.
This tendency of $S(t;N_\mathrm{ag})$ demonstrates that flow-induced scission of micelles with $N_\mathrm{ag}=300$ hardly occurs under this weak extensional flow, which is consistent with $P(N_\mathrm{ag})$ shown in Fig.~\ref{fig:pdf_nag}(a).
For $\mathrm{Wi}\gtrsim 1.6$, while $S(t;N_\mathrm{ag})$ still follows an exponential function, $S(t;N_\mathrm{ag})$ decays faster with increasing $\mathrm{Wi}$.
Thus, $S(t;N_\mathrm{ag})$ provides evidence that individual micelles undergo flow-induced scission under strong extensional flows with $\mathrm{Wi}\gtrsim 2$.

To characterize $\mathrm{Wi}$ dependence of flow-induced scission for each $N_\mathrm{ag}$, we evaluate the average lifetime $\tau_b(N_\mathrm{ag})$ of micelles by fitting $S(t;N_\mathrm{ag})$ to $C\exp\{-t/\tau_b(N_\mathrm{ag})\}$.
Figure~\ref{fig:lifetime}(b) shows $\tau_b(N_\mathrm{ag})$ of micelles with $N_\mathrm{ag}=150$, $200$, and $300$ as a function of $\mathrm{Wi}$ for various $k_BT$ and $\phi$.
Here, $\tau_b(N_\mathrm{ag})$ is normalized by the value $\tau_{b,\mathrm{eq}}(N_\mathrm{ag})$ at equilibrium.
We find that $\tau_b(N_\mathrm{ag})/\tau_{b,\mathrm{eq}}(N_\mathrm{ag})$ almost collapses onto a single function of $\mathrm{Wi}$, which is consistent with the results of flow-induced scission under shear flow~\cite{Koide2022-bp}.
For $\mathrm{Wi}\lesssim 2$, $\tau_b(N_\mathrm{ag})\simeq \tau_{b,\mathrm{eq}}(N_\mathrm{ag})$ holds, whereas for $\mathrm{Wi}\gtrsim 2$, $\tau_b(N_\mathrm{ag})$ decreases with increasing $\mathrm{Wi}$ due to flow-induced scission.
This flow-induced scission of individual micelles for large $\mathrm{Wi}$ causes the decrease in $P(N_\mathrm{ag})$ at large $N_\mathrm{ag}$~[Fig.~\ref{fig:pdf_nag}(a)], thus leading to the decrease in $\eta_E(\dot{\epsilon})$.
As we will show in Sec.~\ref{subsec:discussion_scission}, flow-induced scission not only reduces the fraction of large micelles but also affects micellar dynamics, which is one of the key factors determining $\eta_E(\dot{\epsilon})$.
It should be noted that although the degree of flow-induced scission is well characterized by $\mathrm{Wi}$~(Fig.~\ref{fig:lifetime}), the underlying physical mechanism is nontrivial and thus left for future work.

% %   --------------------
%   --------------------

%   --------------------
\subsection{Stretching and alignment of micelles}

\begin{figure*}
  \centering
      \begin{tabular}{c}
      \begin{minipage}{0.5\hsize}
          \begin{overpic}[width=1\linewidth]{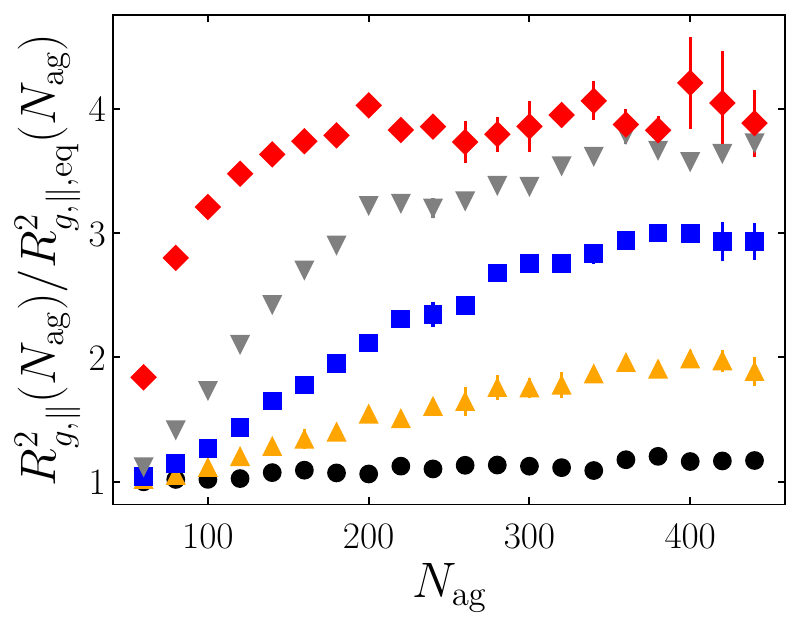}
              \linethickness{3pt}
        \put(0,78){(a)}

          \end{overpic}
      \end{minipage}
      \begin{minipage}{0.5\hsize}
          \begin{overpic}[width=1\linewidth]{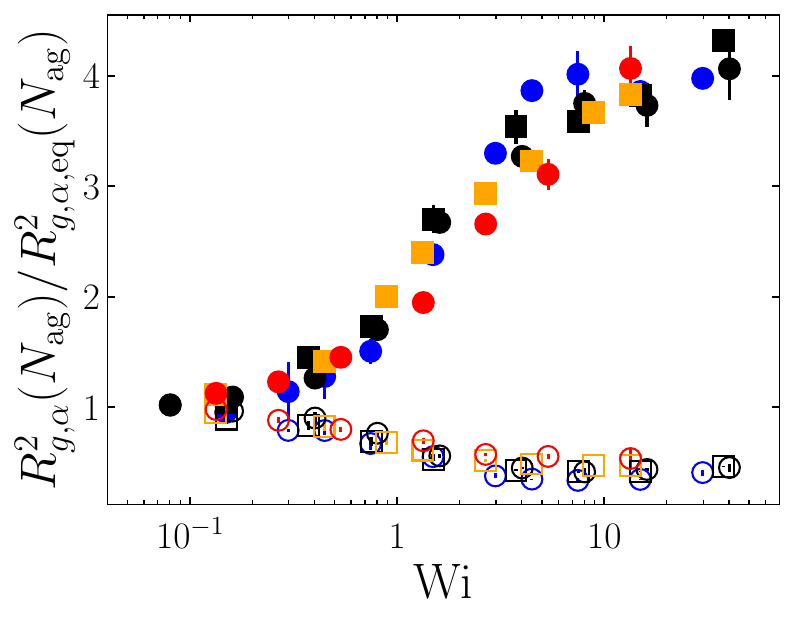}
              \linethickness{3pt}
          \put(0,78){(b)}

          \end{overpic}
      \end{minipage}
      \end{tabular}
      \caption{(a) Mean-square gyration radius $R_{g,\parallel}^2(N_\mathrm{ag})$ of micelles in the extensional direction normalized by the value $R_{g,\parallel,\mathrm{eq}}^2(N_\mathrm{ag})$ at equilibrium as a function of $N_\mathrm{ag}$ for $\mathrm{Wi}=0.16$~(black circle), $0.8$~(orange triangle), $1.6$~(blue square), $4$~(gray inverted triangle), and $16$~(red diamond). (b) $R_{g,\alpha}^2(N_\mathrm{ag})/R_{g,\alpha,\mathrm{eq}}^2(N_\mathrm{ag})$ of micelles with $N_\mathrm{ag}=300$ as a function of $\mathrm{Wi}$. Filled and open symbols denote $R_{g,\parallel}^2(N_\mathrm{ag})$ and $R_{g,\perp}^2(N_\mathrm{ag})$, respectively. In (b), the colors and symbols are the same as in Fig.~\ref{fig:vis}. The error bars denote the standard deviations from three independent simulations.}
      \label{fig:gyration}
\end{figure*}

In this subsection, we investigate the stretching and alignment of micelles to reveal the increasing tendency of the extensional viscosity.
Considering the polydispersity of micellar sizes~(Fig.~\ref{fig:pdf_nag}), we use conditional statistics based on $N_\mathrm{ag}$ to evaluate the stretching and alignment of individual micelles, just as in the case of $\tau_b(N_\mathrm{ag})$.
Specifically, we focus on the mean-square gyration radius in the parallel ($R_{g,\parallel}^2$) and perpendicular ($R_{g,\perp}^2$) directions to the extensional direction~(i.e., the $x$ direction), defined as
\begin{equation}
  R_{g,\parallel}^2(N_\mathrm{ag}) = \left\langle \frac{1}{N_\mathrm{sur}}\sum_{i=1}^{N_{\mathrm{sur}}} (r_{i,x}-r_{G,x})^2\right\rangle_{N_\mathrm{ag}},
\end{equation}
\begin{equation}
  R_{g,\perp}^2(N_\mathrm{ag}) = \frac{1}{2}\left\langle \frac{1}{N_\mathrm{sur}}\sum_{i=1}^{N_{\mathrm{sur}}} \{(r_{i,y}-r_{G,y})^2+(r_{i,z}-r_{G,z})^2\}\right\rangle_{N_\mathrm{ag}},
\end{equation}
where $N_\mathrm{sur}(=3N_\mathrm{ag})$ is the number of surfactant particles in micelles, $\bm{r}_i$ is the position of the $i$-th particle, and $\bm{r}_G$ is the position of the center of mass of micelles. 
Here, $\langle\cdot\rangle_{N_\mathrm{ag}}$ denotes the ensemble average over micelles whose aggregation numbers are within $[N_\mathrm{ag}-\Delta N_\mathrm{ag}/2,N_\mathrm{ag}+\Delta N_\mathrm{ag}/2]$.
Figure~\ref{fig:gyration}(a) shows $R_{g,\parallel}^2(N_\mathrm{ag})$ as a function of $N_\mathrm{ag}$ for various $\mathrm{Wi}$.
Note that $R_{g,\parallel}^2(N_\mathrm{ag})$ is normalized by the value $R_{g,\parallel,\mathrm{eq}}^2(N_\mathrm{ag})$ at equilibrium for each $N_\mathrm{ag}$.
For $\mathrm{Wi}=0.16$, $R_{g,\parallel}^2(N_\mathrm{ag})\simeq R_{g,\parallel,\mathrm{eq}}^2(N_\mathrm{ag})$ holds regardless of $N_\mathrm{ag}$, indicating that micellar structures remain unaffected by extensional flows.
As $\mathrm{Wi}$ increases, $R_{g,\parallel}^2(N_\mathrm{ag})/R_{g,\parallel,\mathrm{eq}}^2(N_\mathrm{ag})$ increases due to micellar stretching and alignment in the extensional direction. 
It is worth emphasizing that $R_{g,\parallel}^2(N_\mathrm{ag})/R_{g,\parallel,\mathrm{eq}}^2(N_\mathrm{ag})$ exhibits two regimes regarding $N_\mathrm{ag}$ dependence.
For small $N_\mathrm{ag}$, $R_{g,\parallel}^2(N_\mathrm{ag})/R_{g,\parallel,\mathrm{eq}}^2(N_\mathrm{ag})$ monotonically increases with $N_\mathrm{ag}$.
This monotonic increase indicates that for fixed $\dot{\epsilon}$, larger micelles are more susceptible to the effect of extensional flows due to their longer relaxation time.
In contrast, $R_{g,\parallel}^2(N_\mathrm{ag})/R_{g,\parallel,\mathrm{eq}}^2(N_\mathrm{ag})$ exhibits a plateau for large $N_\mathrm{ag}$, implying that these large micelles are effectively characterized by the same relaxation time.
Similar saturation of the size dependence was reported in a previous study which investigated the bead-spring model with scission and recombination under shear flow~\cite{Huang2008-sg}.
We also find that for $\mathrm{Wi}=4$ and $16$, the qualitative behavior of $R_{g,\parallel}^2(N_\mathrm{ag})/R_{g,\parallel,\mathrm{eq}}^2(N_\mathrm{ag})$ remains unchanged, but the saturation occurs at smaller values of $N_\mathrm{ag}$ with increasing $\mathrm{Wi}$.
In the next section, we will discuss the physical mechanism underlying this saturation of $R_{g,\parallel}^2(N_\mathrm{ag})/R_{g,\parallel,\mathrm{eq}}^2(N_\mathrm{ag})$ by considering the coupling between micellar scission and dynamics.

We then focus on $\mathrm{Wi}$ dependence of micellar stretching and alignment for fixed $N_\mathrm{ag}$.
Figure~\ref{fig:gyration}(b) shows $R_{g,\parallel}^2(N_\mathrm{ag})/R_{g,\parallel,\mathrm{eq}}^2(N_\mathrm{ag})$ as a function of $\mathrm{Wi}$ for $N_\mathrm{ag}=300$.
We observe that $R_{g,\parallel}^2(N_\mathrm{ag})$ increases with $\mathrm{Wi}$ within the considered range of $\mathrm{Wi}$.
For a fixed $N_\mathrm{ag}$, micelles are more stretched and aligned in the extensional direction as $\mathrm{Wi}$ increases.
The stretching and alignment of individual micelles contribute to the increase in $\eta_E(\dot{\epsilon})$ for $\mathrm{Wi}\lesssim 2$ in a similar way to that observed for polymer models.~\cite{bird1987dynamics}
We also confirm that $R_{g,\parallel}^2(N_\mathrm{ag})/R_{g,\parallel,\mathrm{eq}}^2(N_\mathrm{ag})$ is almost independent of $k_BT$ and $\phi$ for fixed $\mathrm{Wi}$. 
Because the stretching behavior of individual micelles is almost identical in the considered systems, the aggregation-number distribution and micellar kinetics are responsible for $k_BT$ and $\phi$ dependence of $\eta_E(\dot{\epsilon})$ shown in Fig.~\ref{fig:vis}.
We provide a quantitative discussion on this issue in the next section.
Figure~\ref{fig:gyration}(b) also presents the mean-square gyration radius $R_{g,\perp}^2(N_\mathrm{ag})$ in the compression direction, which exhibits a decreasing trend with increasing $\mathrm{Wi}$, contrary to $R_{g,\parallel}^2(N_\mathrm{ag})$.

% %   --------------------
%   --------------------

\begin{figure*}
  \centering
      \begin{tabular}{c}
      \begin{minipage}{0.5\hsize}
          \begin{overpic}[width=1\linewidth]{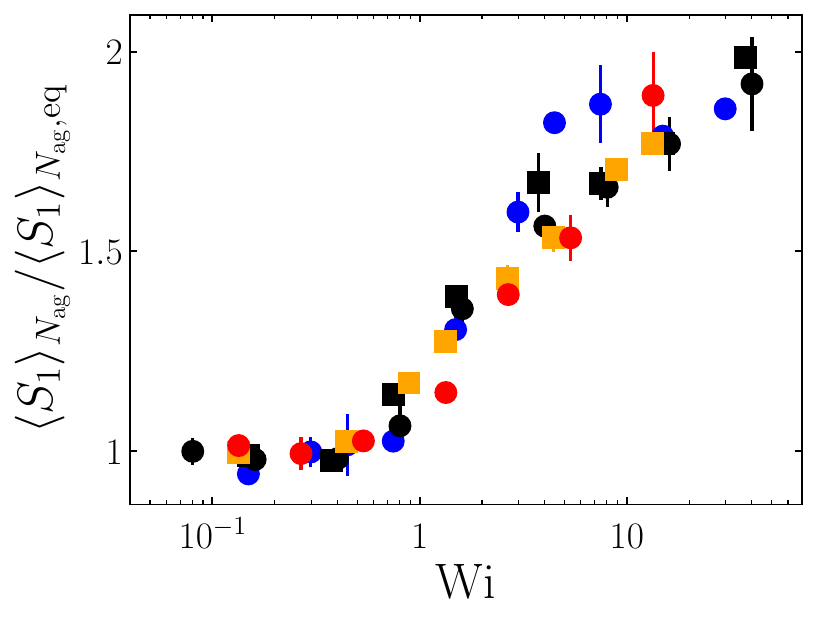}
              \linethickness{3pt}
            \put(0,70){(a)}

          \end{overpic}
      \end{minipage}
      \begin{minipage}{0.5\hsize}
          \begin{overpic}[width=1\linewidth]{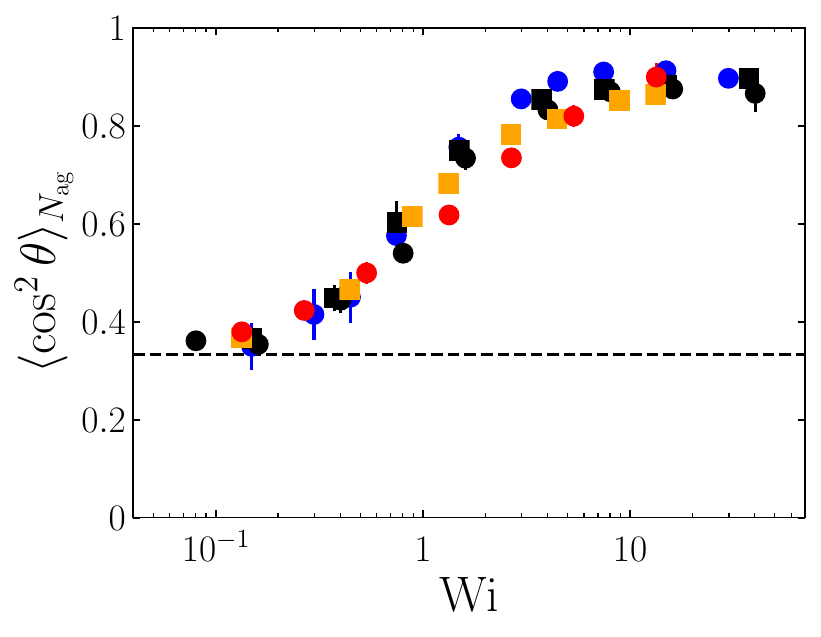}
              \linethickness{3pt}
               \put(0,70){(b)}

          \end{overpic}
      \end{minipage}
      \end{tabular}
      \caption{(a) Average largest eigenvalue $\langle S_1\rangle_{N_\mathrm{ag}}$ of the gyration tensor $\bm{S}$ normalized by the value $\langle S_{1}\rangle_{N_\mathrm{ag},\mathrm{eq}}$ at equilibrium and (b) mean-square cosine $\langle \cos^2\theta\rangle_{N_\mathrm{ag}}$ of the angle $\theta$ between the $x$ axis and the eigenvector $\bm{d}_1$ of $\bm{S}$ corresponding to $S_1$ as functions of the Weissenberg number $\mathrm{Wi}$. Both figures present data for micelles with $N_\mathrm{ag}=300$. The colors and symbols are the same as in Fig.~\ref{fig:vis}. The black dashed line in (b) indicates $\langle \cos^2\theta\rangle_{N_\mathrm{ag}}=1/3$, which corresponds to the isotropic case. The error bars denote the standard deviations from three independent simulations.}
      \label{fig:eigenvalue}
\end{figure*}

Before closing this subsection, we also investigate the eigenvalues and eigenvectors of the instantaneous gyration tensor to evaluate the stretching and alignment of micelles separately.
The instantaneous gyration tensor $\bm{S}$ of individual micelles is defined as 
\begin{equation}
  S_{ij}= \frac{1}{N_\mathrm{sur}}\sum_{k=1}^{N_\mathrm{sur}}(r_{k,i}-r_{G,i})(r_{k,j}-r_{G,j}).
\end{equation}
Let us denote the eigenvalues of $\bm{S}$ by $S_i$~($S_1\geq S_2\geq S_3$) and the corresponding eigenvectors by ${\bm{d}}_i$.
These quantities provide information on the shape and orientation of a particular micelle~\cite{Doi1974-bq,Theodorou1985-il}.
Figure~\ref{fig:eigenvalue}(a) shows the average largest eigenvalue $\langle S_1\rangle_{N_\mathrm{ag}}$ normalized by the value $\langle S_{1}\rangle_{N_\mathrm{ag},\mathrm{eq}}$ at equilibrium as a function of $\mathrm{Wi}$.
To evaluate micellar alignment in the extensional direction, we also show $\langle \cos^2\theta\rangle_{N_\mathrm{ag}}$ as a function of $\mathrm{Wi}$ in Fig.~\ref{fig:eigenvalue}(b), where $\theta$ is the angle between ${\bm{d}}_1$ and the $x$ axis.
We find that both $\langle S_1\rangle_{N_\mathrm{ag}}$ and $\langle \cos^2\theta\rangle_{N_\mathrm{ag}}$ increase with $\mathrm{Wi}$.
This increasing tendency of $\langle S_1\rangle_{N_\mathrm{ag}}$ and $\langle \cos^2\theta\rangle_{N_\mathrm{ag}}$ supports the conclusion that wormlike micelles considered in this study are aligned and stretched by extensional flows, thereby contributing to the increase in $\eta_E(\dot{\epsilon})$.
% %   --------------------
%   --------------------

\section{Discussion}
\label{sec:discussion}

The previous section qualitatively demonstrated that the competition between stretching and flow-induced scission of micelles causes the nonmonotonicity of the extensional viscosity as a function of the extension rate.
In this section, we offer a unified and quantitative description of the extensional viscosity based on micellar properties.
\subsection{Effect of polydispersity on the extensional viscosity}

We aim to understand the extensional viscosity $\eta_E(\dot{\epsilon})$ of wormlike micellar solutions by relying on the theoretical relation for the Rouse-type model, which is a single-chain model consisting of beads interacting with each other via interaction potentials.
Uneyama~\cite{uneyama2025radius} theoretically derived the relation between the viscosity and the gyration radius for the Rouse-type model, starting from the overdamped Langevin equation.
For $M$ polymer chains each of which consists of $N_\mathrm{p}$ beads, $\eta_E(\dot{\epsilon})$ and $R_{g,\alpha}^2$ are related as
\begin{equation}
  \eta_E(\dot{\epsilon})=\rho_\mathrm{p}\zeta \left[R_{g,\parallel}^2+\frac{1}{2}R_{g,\perp}^2\right],\label{eq:vis_mono_gyration}
\end{equation}
where $\rho_\mathrm{p}=MN_\mathrm{p}/V$ is the number density of beads, $\zeta$ is the friction coefficient, and ${R_{g,\alpha}^2}$ is the mean-square gyration radius in the $\alpha$ direction.
Since micellar solutions considered in this study exhibit a broad distribution of $N_\mathrm{ag}$, it is necessary to incorporate polydispersity into Eq.~\eqref{eq:vis_mono_gyration}.
By generalizing the derivation by Uneyama~\cite{uneyama2025radius} to a polydisperse system, we obtain
\begin{equation}
  \eta_E(\dot{\epsilon})=\rho_\mathrm{p}\zeta \left[\langle{R_{g,\parallel}^2}\rangle_\mathrm{w}+\frac{1}{2}\langle{R_{g,\perp}^2}\rangle_\mathrm{w}\right],\label{eq:vis_poly_gyration}
\end{equation}
where $\langle{R_{g,\alpha}^2}\rangle_\mathrm{w}$ is the weighted average of the gyration radius, defined as
\begin{equation}
  \langle{R_{g,\alpha}^2}\rangle_\mathrm{w} = \frac{\sum_{j=1}^M N_{\mathrm{p},j} {R_{g,\alpha,j}^2} }{\sum_{j=1}^M N_{\mathrm{p},j} }.\label{eq:weighted_gyration}
\end{equation}
Here, $N_{\mathrm{p},j}$ is the number of beads in the $j$-th chain, and ${R_{g,\alpha,j}^2}$ denotes the mean-square gyration radius of the $j$-th chain in the $\alpha$ direction.

\begin{figure}
  \centering
  \begin{overpic}[width=0.5\linewidth]{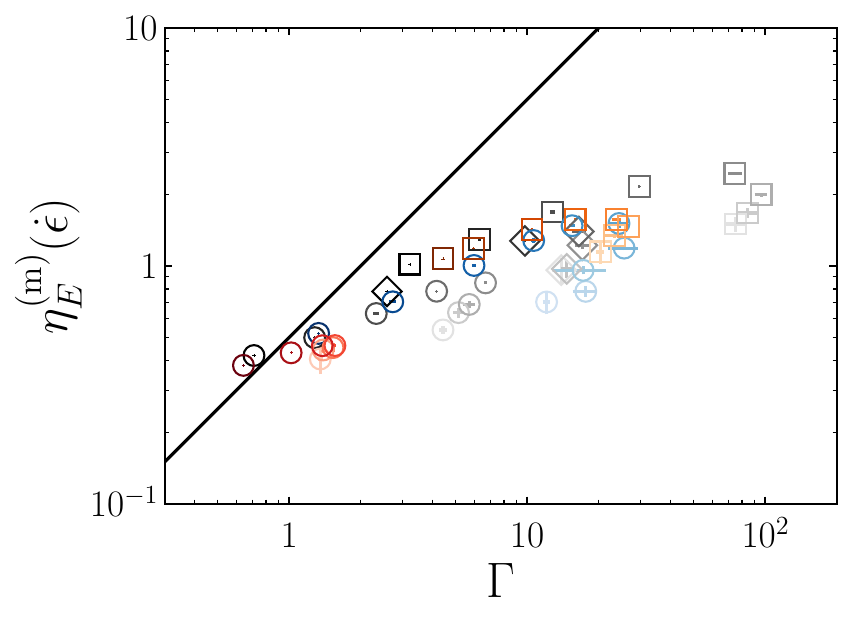} 
  \end{overpic}
  \caption{Micellar contribution $\eta_E^\mathrm{(m)}(\dot{\epsilon})$ to $\eta_E(\dot{\epsilon})$ as a function of $\Gamma$. Different colors denote different values of $k_BT$: blue, $k_BT=0.9$; black, $1$; orange, $1.1$; red, $1.2$. Different symbols denote different values of the surfactant volume fraction $\phi$: circle, $\phi=0.05$; diamond, $0.07$; square, $0.1$. Darker colors correspond to larger values of $\dot{\epsilon}$. The black solid line shows $\eta_E^\mathrm{(m)}(\dot{\epsilon})\propto \Gamma$. The error bars denote the standard deviations from three independent simulations.}
  \label{fig:vis_weighted_gyration}
\end{figure}

Using the relation for the polydisperse Rouse-type chains~[Eq.~\eqref{eq:vis_poly_gyration}], we now consider $\eta_E(\dot{\epsilon})$ of micellar solutions~(Fig.~\ref{fig:vis}).
Instead of $\rho_\mathrm{p}$ in Eq.~\eqref{eq:vis_poly_gyration}, we use the mean number density $\rho_\mathrm{m}$ of surfactant particles comprising micelles.
Considering the one-dimensional growth of wormlike micelles, we assume $N_\mathrm{ag}\propto N_\mathrm{p}$ in Eq.~\eqref{eq:weighted_gyration}.
Accordingly, $\langle{R_{g,\alpha}^2}\rangle_\mathrm{w}$ is calculated as a weighted average with respect to $N_\mathrm{ag}$.
Since $\eta_E(\dot{\epsilon})$ in Eq.~\eqref{eq:vis_poly_gyration} represents the contribution of Rouse-type chains, we focus on the micellar contribution $\eta_E^\mathrm{(m)}(\dot{\epsilon})=\eta_E(\dot{\epsilon})-\eta_{E,\mathrm{w}}$, where $\eta_{E,\mathrm{w}}$ is the water viscosity obtained from additional DPD simulations of water with the same $k_BT$ at a certain $\dot{\epsilon}$.
Depending on $\phi$ and $k_BT$, $\eta_{E,\mathrm{w}}$ constitutes approximately $60\%\text{--}85\%$ of $\eta_E(\dot{\epsilon})$ of surfactant solutions at low $\dot{\epsilon}$.
Consequently, the effect of the polydispersity of micelles is expressed by the relation 
\begin{equation}
  \eta_E^\mathrm{(m)}(\dot{\epsilon}) = \zeta_{\mathrm{m}}\Gamma, \label{eq:vis_gyration_weighted}
\end{equation}
where $\zeta_\mathrm{m}$ is an a priori unknown parameter representing the micellar friction coefficient and $\Gamma=\rho_\mathrm{m}[\langle{R_{g,\parallel}^2}\rangle_\mathrm{w}+\langle{R_{g,\perp}^2}\rangle_\mathrm{w}/2]$.
Figure~\ref{fig:vis_weighted_gyration} shows $\eta_E^\mathrm{(m)}(\dot{\epsilon})$ as a function of $\Gamma$ on a logarithmic scale.
With the assumption that $\zeta_\mathrm{m}$ is constant, Eq.~\eqref{eq:vis_gyration_weighted} predicts that $\eta_E^\mathrm{(m)}(\dot{\epsilon})$ is proportional to $\Gamma$.
However, $\eta_E^\mathrm{(m)}(\dot{\epsilon})\propto \Gamma$ is obviously invalid in our systems, although $\Gamma$ exhibits a weak correlation with $\eta_E^\mathrm{(m)}(\dot{\epsilon})$.
In particular, for fixed $k_BT$ and $\phi$, $\eta_E^\mathrm{(m)}(\dot{\epsilon})$ in the low $\dot{\epsilon}$ regime~(lighter colors) systematically falls below that in the high $\dot{\epsilon}$ regime~(darker colors).
In other words, $\Gamma$ overestimates the contribution from large micelles, which are prevalent in weak extensional flows.
The failure of Eq.~\eqref{eq:vis_gyration_weighted} indicates that polydispersity alone is insufficient to explain the behavior of $\eta_E(\dot{\epsilon})$ in micellar solutions.
Although one may argue that the relation for the Rouse-type model does not inherently hold in DPD simulations, we have confirmed that Eq.~\eqref{eq:vis_mono_gyration} offers a relation between $\eta_E(\dot{\epsilon})$ and $R_{g,\alpha}^2$ for different values of $\phi$, $k_BT$, $N_\mathrm{p}$ and $\dot{\epsilon}$ in the case of polymer solutions~(see Appendix~B).
In the next subsection, we refine the theory by considering the effect of scission on micellar dynamics.
%   --------------------
%   --------------------

\subsection{\label{subsec:discussion_scission}Effect of micellar scission on the extensional viscosity}

This subsection focuses on how micellar scission affects the relation between the extensional viscosity and the gyration radius of micelles.
The relation for the Rouse-type model~[Eq.~\eqref{eq:vis_mono_gyration}] is derived from the equation of motion for a single chain in flows, combined with the ensemble average.
In this derivation, the constituent beads in the chain retain their identity, meaning that the individual particles being tracked never change.
In contrast, micelles constantly undergo scission and recombination, leading to the exchange of constituent surfactants.
Because this effect is absent from Eq.~\eqref{eq:vis_gyration_weighted}, that equation fails to predict the relation between $\eta_E^\mathrm{(m)}(\dot{\epsilon})$ and the gyration radius of micelles.
To rigorously incorporate the effect of kinetics on $\eta_E^\mathrm{(m)}(\dot{\epsilon})$, we may need to modify the original theory by coupling the dynamic equations with scission and recombination kinetics.
Unfortunately, such a modification requires more detailed information on micellar kinetics, and the analysis would be quite complicated.
Instead, we construct a simple yet physically plausible model that incorporates the effect of micellar kinetics. 
Within this model, we introduce a characteristic aggregation number $\widetilde{N}_\Lambda(\dot{\epsilon})$, which is determined by the competition between the timescales of micellar rotational dynamics and scission.
We will demonstrate that $\widetilde{N}_\Lambda(\dot{\epsilon})$ plays a role in distinguishing the manner in which micelles contribute to $\eta_E^\mathrm{(m)}(\dot{\epsilon})$.

\begin{figure*}
  \centering
      \begin{tabular}{c}
      \begin{minipage}{0.5\hsize}
          \begin{overpic}[width=1\linewidth]{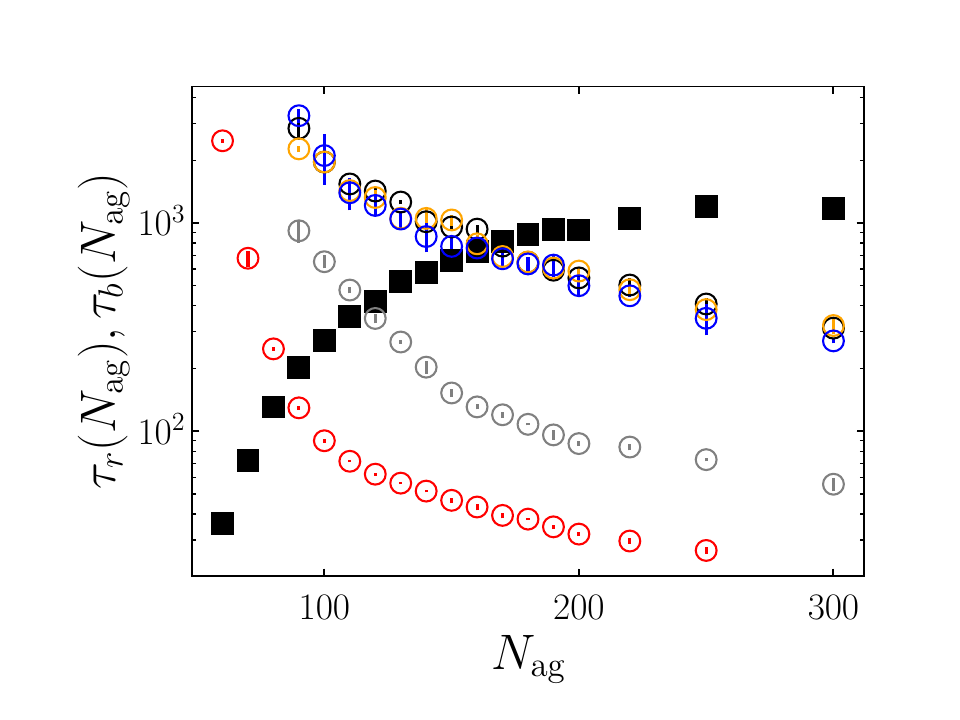}
              \linethickness{3pt}
                    \put(0,70){(a)}

          \end{overpic}
      \end{minipage}
      \begin{minipage}{0.5\hsize}
          \begin{overpic}[width=1\linewidth]{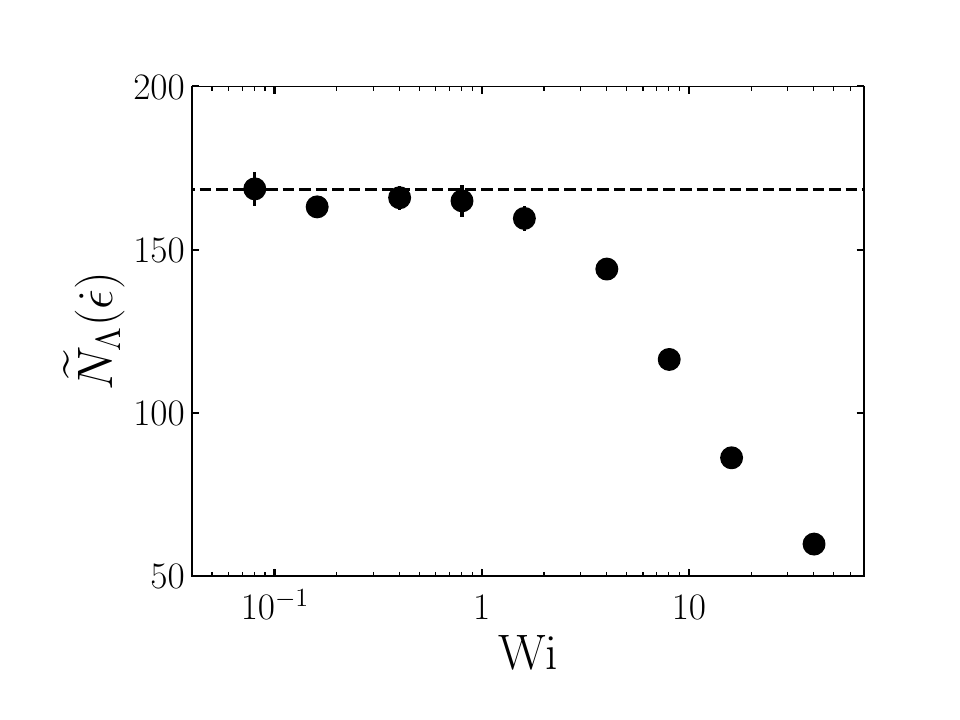}
              \linethickness{3pt}
                 \put(0,70){(b)}

          \end{overpic}
      \end{minipage}
      \end{tabular}
  \caption{(a) Rotational relaxation time $\tau_r(N_\mathrm{ag})$~(filled square) and average lifetime $\tau_b(N_\mathrm{ag})$~(open circle) as functions of the aggregation number $N_\mathrm{ag}$. Different colors denote different values of $\mathrm{Wi}$: black, $\mathrm{Wi}=0$; orange, $0.16$; blue, $1.6$; gray, $8$; red, $16$. (b) Largest dynamically effective size $\widetilde{N}_\Lambda(\dot{\epsilon})$ as a function of $\mathrm{Wi}$. The black dashed line indicates the value $N_\Lambda$ at equilibrium. Both figures present data for $k_BT =1$ and $\phi=0.05$. The error bars denote the standard deviations from three independent simulations.}
    
      \label{fig:longest_relaxation}
\end{figure*}

The key point is that micellar scission imposes the upper bound on the aggregation number that is effective in micellar dynamics, as demonstrated in a previous study~\cite{Koide2022-bp}.
Figure~\ref{fig:longest_relaxation}(a) shows the rotational relaxation time $\tau_r(N_\mathrm{ag})$ of micelles as a function of $N_\mathrm{ag}$.
Following a previous study~\cite{Koide2022-bp}, we evaluate $\tau_r(N_\mathrm{ag})$ using the autocorrelation function of the eigenvector $\bm{d}_1$ of the gyration tensor $\bm{S}$ corresponding to the largest eigenvalue $S_1$.
We observe that $\tau_r(N_\mathrm{ag})$ is a monotonically increasing function of $N_\mathrm{ag}$, where the definition of $\tau_r(N_\mathrm{ag})$ includes no effect of micellar scission.
Figure~\ref{fig:longest_relaxation}(a) also shows $\tau_b(N_\mathrm{ag})$ as a function of $N_\mathrm{ag}$, which is a monotonically decreasing function of $N_\mathrm{ag}$. 
For large $N_\mathrm{ag}$, $\tau_r(N_\mathrm{ag})>\tau_b(N_\mathrm{ag})$ holds.
This inequality indicates that such large micelles are likely to undergo scission during their slow rotational relaxation.
Consequently, the slow relaxation modes for which $\tau_r(N_\mathrm{ag})>\tau_b(N_\mathrm{ag})$ effectively disappear, and the longest relaxation time $\tau_\Lambda$ and the largest dynamically effective size $N_\Lambda$ in the system are bounded by the scission effect.
Previous studies~\cite{Koide2022-bp,Koide2023-ao} estimated $\tau_\Lambda$ and $N_\Lambda$ from the intersection of $\tau_r(N_\mathrm{ag})$ and $\tau_b(N_\mathrm{ag})$.
Then, the longest relaxation time $\tau(N_\mathrm{ag})$ of individual micelles with $N_\mathrm{ag}$ was assumed to follow
\begin{equation}
  \tau(N_\mathrm{ag}) = 
  \begin{cases}
      \tau_r(N_\mathrm{ag}) & (N_\mathrm{ag} < N_\Lambda) \\
      \tau_\Lambda & (N_\mathrm{ag} \geq N_\Lambda) .
  \end{cases}
  \label{eq:longest_relaxation}
\end{equation}
Here, for $N_\mathrm{ag}<N_\Lambda$, $\tau(N_\mathrm{ag})=\tau_r(N_\mathrm{ag})$ holds because scission hardly affects micellar dynamics, whereas all micelles with $N_\mathrm{ag}\geq N_\Lambda$ are characterized by the common timescale $\tau_\Lambda$ due to the limit on the dynamically effective aggregation number.
Figure~\ref{fig:gyration}(a) demonstrates that for fixed $\dot{\epsilon}$, $R_{g,\parallel}^2(N_\mathrm{ag})/R_{g,\parallel,\mathrm{eq}}^2(N_\mathrm{ag})$ initially increases with $N_\mathrm{ag}$ due to the increase in $\tau(N_\mathrm{ag})$.
However, $R_{g,\parallel}^2(N_\mathrm{ag})/R_{g,\parallel,\mathrm{eq}}^2(N_\mathrm{ag})$ saturates above a certain $N_\mathrm{ag}$, which is consistent with the above-mentioned concept.
The present study also demonstrated that $\tau_\Lambda$, which is the longest timescale in the system, characterizes the $\dot{\epsilon}$ dependence of $\eta_E(\dot{\epsilon})$~(Fig.~\ref{fig:vis}).
Previous studies~\cite{Koide2022-bp,Koide2023-ao} also demonstrated the validity of Eq.~\eqref{eq:longest_relaxation} for micellar scission and alignment under shear flow.
Similar concepts regarding the coupling between dynamics and scission of chains were applied to both unentangled and entangled micellar solutions~\cite{Faivre1986-nb,Cates1987-tv,Huang2006-ov,Huang2009-nw}.

We aim to extend the largest dynamically effective size $N_\Lambda$ to its non-equilibrium version $\widetilde{N}_{\Lambda}(\dot{\epsilon})$ under uniaxial extensional flow.
As shown in Fig.~\ref{fig:lifetime}, extensional flows promote micellar scission.
Figure~\ref{fig:longest_relaxation}(a) shows $\tau_b(N_\mathrm{ag})$ as a function of $N_\mathrm{ag}$ for various $\mathrm{Wi}$, illustrating an overall decrease in $\tau_b(N_\mathrm{ag})$ with increasing $\mathrm{Wi}$. 
According to the concept of Eq.~\eqref{eq:longest_relaxation}, the decrease in $\tau_b(N_\mathrm{ag})$ causes the reduction of the largest dynamically effective size.
Here, we estimate $\widetilde{N}_\Lambda(\dot{\epsilon})$ by using the values of $\tau_b(N_\mathrm{ag})$ at each $\dot{\epsilon}$ instead of the equilibrium values $\tau_{b,\mathrm{eq}}(N_\mathrm{ag})$.
Although rotational relaxation behavior may change under strong extensional flow, we focus on the flow effects on micellar scission, which are significant as demonstrated in Fig.~\ref{fig:lifetime}.
Figure~\ref{fig:longest_relaxation}(b) presents $\widetilde{N}_\Lambda(\dot{\epsilon})$ as a function of $\mathrm{Wi}$ for $k_BT=1$ and $\phi=0.05$.
For $\mathrm{Wi}\lesssim 2$, since flow-induced scission hardly occurs, $\widetilde{N}_\Lambda(\dot{\epsilon})$ equals to the value at equilibrium.
For $\mathrm{Wi}\gtrsim 2$, $\widetilde{N}_\Lambda(\dot{\epsilon})$ decreases with increasing $\mathrm{Wi}$ due to the reduction of $\tau_b(N_\mathrm{ag})$ by flow-induced scission.
In fact, Fig.~\ref{fig:gyration}(b) demonstrates that for $\mathrm{Wi}\gtrsim 2$, $R_{g,\parallel}^2(N_\mathrm{ag})/R_{g,\parallel,\mathrm{eq}}^2(N_\mathrm{ag})$ saturates at smaller $N_\mathrm{ag}$ as $\mathrm{Wi}$ increases.
A previous study~\cite{Koide2023-ao} reported such a variation in $\widetilde{N}_\Lambda(\dot{\epsilon})$ in the context of micellar alignment under shear flow.
It is worth noting that since Eq.~\eqref{eq:longest_relaxation} is based on a crude approximation that slower relaxation modes than $\tau_\Lambda$ are completely ignored, $\widetilde{N}_\Lambda(\dot{\epsilon})$ underestimates the saturation point of $R_{g,\parallel}^2(N_\mathrm{ag})/R_{g,\parallel,\mathrm{eq}}^2(N_\mathrm{ag})$.

% %   --------------------
%   --------------------

We consider the effect of the coupling between micellar scission and dynamics on $\eta_E^\mathrm{(m)}(\dot{\epsilon})$ through the largest dynamically effective size $\widetilde{N}_\Lambda(\dot{\epsilon})$.
For $N_\mathrm{ag}\lesssim \widetilde{N}_\Lambda(\dot{\epsilon})$, it is sufficient to consider only the effect of polydispersity because scission hardly prevents the rotational motion of micelles.
Thus, the contribution of micelles with $N_\mathrm{ag}< \widetilde{N}_\Lambda(\dot{\epsilon})$ to $\eta_E^\mathrm{(m)}(\dot{\epsilon})$ is approximately proportional to $\Gamma_{\widetilde{N}_\Lambda(\dot{\epsilon})}^{<}$, defined as
\begin{equation}
  \Gamma_{\widetilde{N}_\Lambda(\dot{\epsilon})}^{<} = \rho_{\widetilde{N}_\Lambda(\dot{\epsilon})}^<\left[{\left\langle R_{g,\parallel}^2\right\rangle}_{\mathrm{w},\widetilde{N}_\Lambda(\dot{\epsilon})}^<+\frac{1}{2}{\left\langle R_{g,\perp}^2\right\rangle}_{\mathrm{w},\widetilde{N}_\Lambda(\dot{\epsilon})}^<\right],\label{eq:gamma_small}
\end{equation}
where $ \rho_{\widetilde{N}_\Lambda(\dot{\epsilon})}^<$ is the mean number density of surfactant particles that belong to micelles with $N_\mathrm{ag}< \widetilde{N}_\Lambda(\dot{\epsilon})$ and ${\langle\cdot\rangle}_{\mathrm{w},{\widetilde{N}_\Lambda}(\dot{\epsilon})}^<$ denotes the weighted average over micelles with $N_\mathrm{ag}<\widetilde{N}_\Lambda(\dot{\epsilon})$.
In contrast, for $N_\mathrm{ag}\geq\widetilde{N}_\Lambda(\dot{\epsilon})$, since $\tau_b(N_\mathrm{ag})\lesssim \tau_r(N_\mathrm{ag})$, the $N_\mathrm{ag}$-dependent characteristics of individual micelles almost disappear.
Over the repeated scission and recombination events, the contribution of micelles with $N_\mathrm{ag}\geq\widetilde{N}_\Lambda(\dot{\epsilon})$ can be approximated as that of monodisperse micelles with $N_\mathrm{ag}=\widetilde{N}_\Lambda(\dot{\epsilon})$, thus being proportional to $\Gamma_{\widetilde{N}_\Lambda(\dot{\epsilon})}^{>}$, defined as
\begin{equation}
  \Gamma_{\widetilde{N}_\Lambda(\dot{\epsilon})}^{>}=\rho_{\widetilde{N}_\Lambda(\dot{\epsilon})}^>\left[R_{g,\parallel}^2(\widetilde{N}_\Lambda(\dot{\epsilon}))+\frac{1}{2}R_{g,\perp}^2(\widetilde{N}_\Lambda(\dot{\epsilon}))\right],\label{eq:gamma_large}
\end{equation}
where $ \rho_{\widetilde{N}_\Lambda(\dot{\epsilon})}^>$ is the mean number density of surfactant particles that belong to micelles with $N_\mathrm{ag}\geq \widetilde{N}_\Lambda(\dot{\epsilon})$.
The presence of the single characteristic scale $\widetilde{N}_\Lambda(\dot{\epsilon})$ arising from micellar kinetics resembles the physical mechanism underlying the single relaxation time observed in entangled wormlike micelles with fast kinetics~\cite{Cates1987-tv,Cates1990-zc}.
With Eqs.~\eqref{eq:gamma_small} and \eqref{eq:gamma_large}, we obtain the relation
\begin{equation}
  \eta_E^\mathrm{(m)}(\dot{\epsilon}) =\zeta_{\mathrm{m}}\left[ \Gamma_{\widetilde{N}_\Lambda(\dot{\epsilon})}^{<}+\Gamma_{\widetilde{N}_\Lambda(\dot{\epsilon})}^{>}\right].\label{eq:vis_gyration}
\end{equation}

\begin{figure*}
  \centering
      \begin{tabular}{c}
      \begin{minipage}{0.5\hsize}
          \begin{overpic}[width=0.9\linewidth]{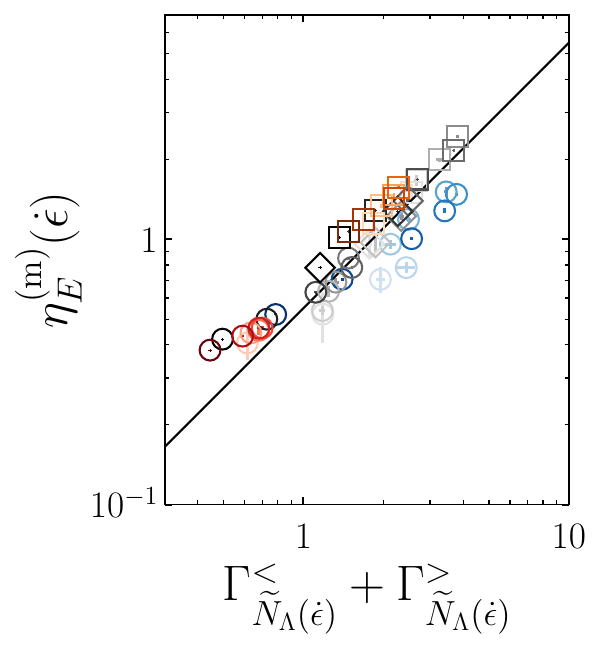}
              \linethickness{3pt}
        \put(5,95){(a)}

          \end{overpic}
      \end{minipage}
      \begin{minipage}{0.5\hsize}
          \begin{overpic}[width=0.9\linewidth]{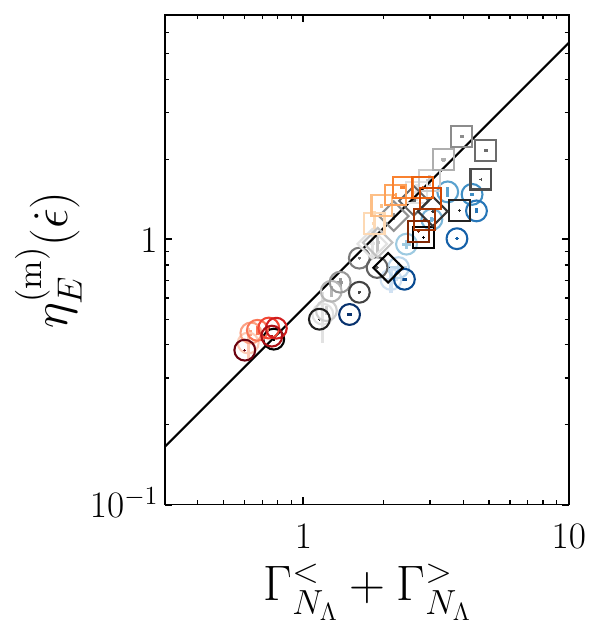}
              \linethickness{3pt}
        \put(5,95){(b)}

          \end{overpic}
      \end{minipage}
      \end{tabular}
      \caption{Micellar contribution $\eta_E^\mathrm{(m)}(\dot{\epsilon})$ to $\eta_E(\dot{\epsilon})$ as a function of (a) $\Gamma_{\widetilde{N}_\Lambda(\dot{\epsilon})}^{<}+\Gamma_{\widetilde{N}_\Lambda(\dot{\epsilon})}^{>}$ and (b) $\Gamma_{N_{\Lambda}}^{<}+\Gamma_{N_{\Lambda}}^{>}$. The colors and symbols are the same as in Fig.~\ref{fig:vis_weighted_gyration}. The black solid lines in (a) and (b) indicate $\eta_E^\mathrm{(m)}(\dot{\epsilon})\propto \Gamma_{\widetilde{N}_\Lambda(\dot{\epsilon})}^{<}+\Gamma_{\widetilde{N}_\Lambda(\dot{\epsilon})}^{>}$ and $\eta_E^\mathrm{(m)}(\dot{\epsilon})\propto \Gamma_{N_\Lambda}^{<}+\Gamma_{N_\Lambda}^{>}$, respectively. The error bars denote the standard deviations from three independent simulations.}

      \label{fig:vis_gyration}
\end{figure*}

To verify the relation~[Eq.~\eqref{eq:vis_gyration}], we show $\eta_E^\mathrm{(m)}(\dot{\epsilon})$ as a function of $\Gamma_{\widetilde{N}_\Lambda(\dot{\epsilon})}^{<}+\Gamma_{\widetilde{N}_\Lambda(\dot{\epsilon})}^{>}$ in Fig.~\ref{fig:vis_gyration}~(a).
We observe that $\eta_E^\mathrm{(m)}(\dot{\epsilon})$ is proportional to $\Gamma_{\widetilde{N}_\Lambda(\dot{\epsilon})}^{<}+\Gamma_{\widetilde{N}_\Lambda(\dot{\epsilon})}^{>}$ and almost collapses onto a single function irrespective of $k_BT$, $\phi$, and $\dot{\epsilon}$, as is evident when compared with Fig.~\ref{fig:vis_weighted_gyration}, where only the polydispersity effect is considered.
Under the assumption that $\zeta_{\mathrm{m}}$ is constant, the observed proportionality confirms that Eq.~\eqref{eq:vis_gyration} successfully incorporates the effect of micellar scission.
To demonstrate the relevance of $\widetilde{N}_\Lambda(\dot{\epsilon})$ in $\eta_E^\mathrm{(m)}(\dot{\epsilon})$, Fig.~\ref{fig:vis_gyration}(b) shows $\eta_E^\mathrm{(m)}(\dot{\epsilon})$ as a function of $\Gamma_{N_\Lambda}^{<}+\Gamma_{N_\Lambda}^{>}$, which is based on the equilibrium value $N_{\Lambda}$ of the largest dynamically effective size instead of $\widetilde{N}_\Lambda(\dot{\epsilon})$.
Although $\Gamma_{N_\Lambda}^{<}+\Gamma_{N_\Lambda}^{>}$ exhibits a proportional relationship with $\eta_E^\mathrm{(m)}(\dot{\epsilon})$, scattering of data is observed when varying $\dot{\epsilon}$ for fixed $\phi$ and $k_BT$.
More concretely, $\Gamma_{N_\Lambda}^{<}+\Gamma_{N_\Lambda}^{>}$ appears to be overestimated at high $\dot{\epsilon}$.
Thus, the variation in $\widetilde{N}_\Lambda(\dot{\epsilon})$ due to flow-induced scission plays a crucial role in $\eta_E^\mathrm{(m)}(\dot{\epsilon})$.

\begin{figure}
  \centering
  \begin{overpic}[width=0.5\linewidth]{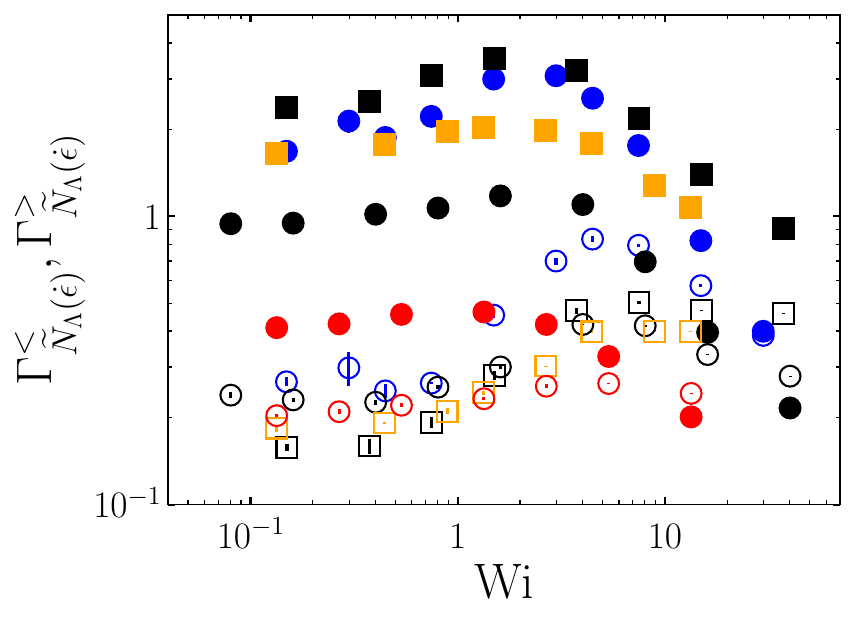} 
  \end{overpic}
  \caption{$\Gamma_{\widetilde{N}_\Lambda(\dot{\epsilon})}^{<}$~(open symbols) and $\Gamma_{\widetilde{N}_\Lambda(\dot{\epsilon})}^{>}$~(filled symbols) as functions of the Weissenberg number $\mathrm{Wi}$. The colors and symbols are the same as in Fig.~\ref{fig:vis}. The error bars denote the standard deviations from three independent simulations.}
  \label{fig:gamma_contribution}
\end{figure}

The respective contributions from $\Gamma_{\widetilde{N}_\Lambda(\dot{\epsilon})}^{<}$ and $\Gamma_{\widetilde{N}_\Lambda(\dot{\epsilon})}^{>}$ also provide insight into the extensional properties of the systems considered.
Figure~\ref{fig:gamma_contribution} shows $\Gamma_{\widetilde{N}_\Lambda(\dot{\epsilon})}^{<}$~(open symbols) and $\Gamma_{\widetilde{N}_\Lambda(\dot{\epsilon})}^{>}$~(filled symbols) as functions of $\mathrm{Wi}$.
Except for large $\mathrm{Wi}$, $\Gamma_{\widetilde{N}_\Lambda(\dot{\epsilon})}^{>}$ is larger than $\Gamma_{\widetilde{N}_\Lambda(\dot{\epsilon})}^{<}$, indicating that micelles with $N_\mathrm{ag}$ larger than $\widetilde{N}_\Lambda(\dot{\epsilon})$ mainly contribute to $\eta_E^\mathrm{(m)}(\dot{\epsilon})$.
The dominant contribution of $\Gamma_{\widetilde{N}_\Lambda(\dot{\epsilon})}^{>}$ qualitatively explains why a single characteristic timescale $\tau_\Lambda$ is sufficient to describe $\eta_E(\dot{\epsilon})$ of surfactant solutions~(Fig.~\ref{fig:vis}).
In addition, $\Gamma_{\widetilde{N}_\Lambda(\dot{\epsilon})}^{>}$ exhibits a nonmonotonic dependence on $\mathrm{Wi}$, with a peak observed around $\mathrm{Wi}\simeq 2$.
Therefore, on the basis of Eq.~\eqref{eq:gamma_large}, the nonmonotonic behavior of $\eta_E(\dot{\epsilon})$ around $\mathrm{Wi}\simeq 2$ is caused by the competition between an increase in $R_{g,\parallel}^2(\widetilde{N}_\Lambda(\dot{\epsilon}))$ due to micellar stretching and decreases in $\rho_{\widetilde{N}_\Lambda(\dot{\epsilon})}^>$ and $\widetilde{N}_\Lambda(\dot{\epsilon})$ due to flow-induced scission.

In summary, the extensional viscosity of wormlike micellar solutions is mainly dominated by three factors: the gyration radius of micelles~(especially in the extensional direction), the aggregation-number distribution, and the largest dynamically effective size of micelles.
Although our relation~[Eq.~\eqref{eq:vis_gyration}] explains the behavior of $\eta_E(\dot{\epsilon})$ for different $k_BT$, $\phi$, and $\dot{\epsilon}$, the collapse is still less clear compared with polymer solutions~(Fig.~\ref{fig:vis_gyration_polymer}).
This discrepancy is mainly attributed to the bold approximations of the effect of micellar kinetics in Eqs.~\eqref{eq:longest_relaxation}, \eqref{eq:gamma_small}, and \eqref{eq:gamma_large}.
Thus, deriving a rigorous relation for the Rouse-type model with scission and recombination kinetics remains an important issue for future research.
Moreover, the physical origin of the value of $\mathrm{\zeta}_\mathrm{m}$ in Eq.~\eqref{eq:vis_gyration} warrants further investigation.

Finally, it is worth referring to the connection between experiments and DPD simulations, although our DPD model does not explicitly assume specific surfactant species.
Eq.~\eqref{eq:vis_gyration} provides insight into the large value of $\eta_E(\dot{\epsilon})/3\eta_0~(\gtrsim 10)$ observed in experiments.~\cite{Walker1996-cf,Lu1997-vi,Lin2001-ge,Rothstein2003-op,Chellamuthu2008-sm,Tamano2017-og,Omidvar2019-rz}
Eq.~\eqref{eq:vis_gyration} indicates that to reproduce such a large value of $\eta_E(\dot{\epsilon})/3\eta_0$ with DPD simulations, it is necessary to increase $\widetilde{N}_\Lambda(\dot{\epsilon})$ by simulating long-lived micelles with large values of $\tau_b(N_\mathrm{ag})$.
For this purpose, one can enhance the hydrophobicity and change the topology of surfactant models.
Since DPD simulations of larger micelles with longer lifetimes require a larger simulation box and longer computational times, such analysis is left for future work.

% %   --------------------
%   --------------------

%   --------------------
%   --------------------

\section{Conclusion}

We have examined the steady-state extensional viscosity of wormlike micellar solutions by conducting DPD simulations with the GKR method~(Fig.~\ref{fig:snapshot}).
One of the most important conclusions is that the steady-state extensional viscosity $\eta_E(\dot{\epsilon})$ exhibits a nonmonotonic dependence on the extension rate $\dot{\epsilon}$, as observed in previous experiments~\cite{Prudhomme1994-tu,Hu1994-dj,Walker1996-cf,Chen1997-xd,Fischer1997-ji,Lin2001-ge}.
Our detailed analyses have demonstrated that this nonmonotonic behavior of $\eta_E(\dot{\epsilon})$ results from the competition between micellar stretching and scission by extensional flows.
Specifically, we have revealed that flow-induced scission of micelles occurs for $\mathrm{Wi}\gtrsim 2$ in the systems considered by evaluating the average lifetime $\tau_b(N_\mathrm{ag})$ of micelles for each $N_\mathrm{ag}$~(Fig.~\ref{fig:lifetime}).
Consequently, the population of large micelles decreases~(Fig.~\ref{fig:pdf_nag}), leading to a decrease in the mean aggregation number $\overline{N}_\mathrm{ag}$~(Fig.~\ref{fig:mean_nag}).
In contrast, the mean-square gyration radius $R_{g,\parallel}^2(N_\mathrm{ag})$ of micelles in the extensional direction increases with $\mathrm{Wi}$, indicating micellar stretching by strong flows~(Fig.~\ref{fig:gyration}).
Therefore, we conclude that for $\mathrm{Wi}\lesssim 2$, micellar stretching causes an increase in $\eta_E(\dot{\epsilon})$, whereas for $\mathrm{Wi}\gtrsim 2$, flow-induced scission becomes significant and decreases $\eta_E(\dot{\epsilon})$ by surpassing the contribution from micellar stretching.

We have also provided a quantitative relation between $\eta_E(\dot{\epsilon})$ and micellar properties to understand the dependence of $\eta_E(\dot{\epsilon})$ on $k_BT$, $\phi$, and $\dot{\epsilon}$ in a unified manner.
We have extended the relation between $\eta_E(\dot{\epsilon})$ and the gyration radius for the Rouse-type model~\cite{uneyama2025radius} by incorporating the specific properties of surfactant micelles, i.e., the aggregation number distribution $P(N_\mathrm{ag})$ and the effect of scission on micellar dynamics.
We have shown that the polydispersity of $N_\mathrm{ag}$ alone is insufficient to explain the complicated behavior of the micellar contribution $\eta_E^\mathrm{(m)}(\dot{\epsilon})$ to $\eta_E(\dot{\epsilon})$~(Fig.~\ref{fig:vis_weighted_gyration}) because micellar scission not only changes $P(N_\mathrm{ag})$ but also restricts the dynamically effective size of micelles.
To consider the effect of scission on micellar dynamics, we have extended the largest dynamically effective size $N_\Lambda$, which was proposed in a previous study for equilibrium cases~\cite{Koide2022-bp}, to its non-equilibrium version $\widetilde{N}_\Lambda(\dot{\epsilon})$ under uniaxial extensional flow.
Incorporating this scission effect into the theoretical relation for the Rouse-type model has allowed us to quantitatively relate $\eta_E^\mathrm{(m)}(\dot{\epsilon})$ to micellar properties~[Eq.~\eqref{eq:vis_gyration}].
The proposed relation provides a unified description of $\dot{\epsilon}$ dependence of $\eta_E(\dot{\epsilon})$ for different $k_BT$ and $\phi$~(Fig.~\ref{fig:vis_gyration}).
Therefore, we have elucidated the dominant factors of the extensional viscosity of unentangled wormlike micellar solutions: the gyration radius in the extensional direction, the distribution of aggregation numbers, and the largest dynamically effective size of micelles.

In the present study, we have paved the way for understanding the relation between the extensional rheology of wormlike micellar solutions and the micellar structures and kinetics.
Since the scission properties of micelles, which have been shown to play a crucial role in $\eta_E(\dot{\epsilon})$, may depend on the flow kinematics, it would be an important future study to apply our approach to biaxial and planar extensional flows.
In addition, the degree of flow-induced scission may vary depending on the hydrophobicity and topology of surfactants, as well as the counterion conditions in the case of ionic surfactants.
Thus, it is also essential to explore various systems consisting of different types of surfactants, which is left for future study.

\begin{acknowledgement}

This work was supported by JSPS Grants-in-Aid for Scientific Research (21J21061 and 24KJ0109) and JST, ACT-X (JPMJAX24D5). 
The DPD simulations were mainly conducted using the General Projects on supercomputer ``Flow'' at Information Technology Center, Nagoya University.
A part of the simulations was conducted under the auspices of the National Institute for Fusion Science Collaboration Research Programs (NIFS22KISS010) and  the JAXA Supercomputer System Generation 3 (JSS3).

\end{acknowledgement}
\section*{\label{sec:Appendix_A}Appendix A: Temeperature control under uniaxial extensional flow}
\renewcommand{\theequation}{A\arabic{equation} }
\setcounter{equation}{0}

\begin{figure}
  \centering
  \begin{overpic}[width=0.5\linewidth]{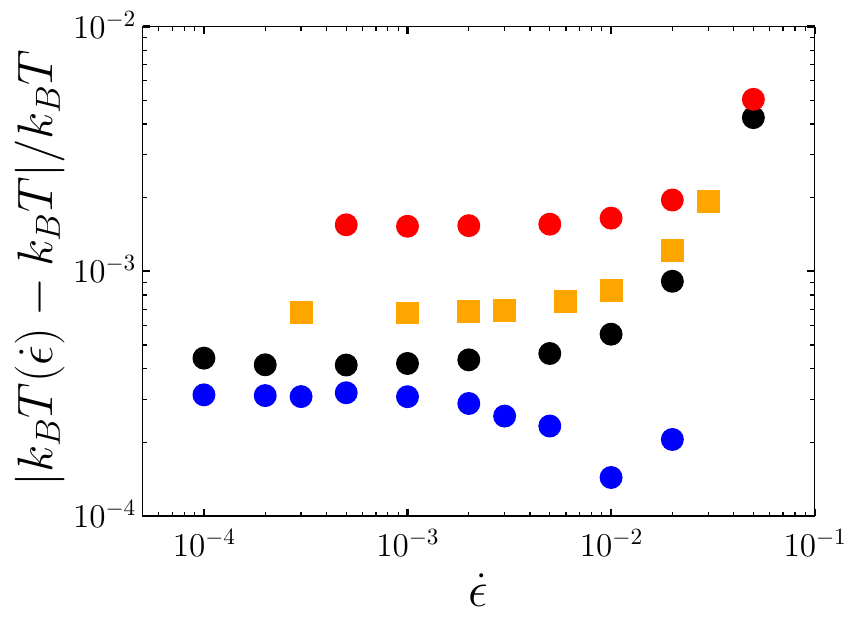} 
  \end{overpic}
  \caption{Relative temperature error $|k_BT(\dot{\epsilon})-k_BT|/k_BT$ as a function of the extension rate $\dot{\epsilon}$ for $(k_BT,\phi)=(0.9,0.05)$~(blue circle), $(1,0.05)$~(black circle), $(1.1,0.1)$~(orange square), and $(1.2,0.05)$~(red circle).}
  \label{fig:temp}
\end{figure}

This appendix provides details on the accuracy of temperature control under uniaxial extensional flow.
Figure~\ref{fig:temp} shows the relative temperature error $|k_BT(\dot{\epsilon})-k_BT|/k_BT$ as a function of the extension rate $\dot{\epsilon}$ for various values of $k_BT$.
Here, $k_BT(\dot{\epsilon})$ is evaluated as 
\begin{equation}
  k_BT(\dot{\epsilon}) = \frac{1}{3}\langle \bm{p}^2\rangle,
\end{equation}
where the particle mass $m$ is set to unity and $\bm{p}$ is the peculiar momentum of DPD particles.
We confirm that $|k_BT(\dot{\epsilon})-k_BT|/k_BT$ remains below $10^{-2}$ within the range considered, although $T(\dot{\epsilon})$ increases with $\dot{\epsilon}$ regardless of the value of $k_BT$, as reported in extensional flow simulations of the Kremer--Grest model using the Langevin thermostat~\cite{Murashima2018-te}.
Note that in the case of $k_BT=0.9$, $k_BT(\dot{\epsilon})-k_BT$ takes negative values for $\dot{\epsilon}\leq 10^{-2}$.
%   --------------------
%   --------------------

\section*{\label{sec:Appendix_B}Appendix B: Relation between the extensional viscosity and the gyration radius for polymers}
\renewcommand{\theequation}{B\arabic{equation} }

\begin{figure}
  \centering
  \begin{overpic}[width=0.5\linewidth]{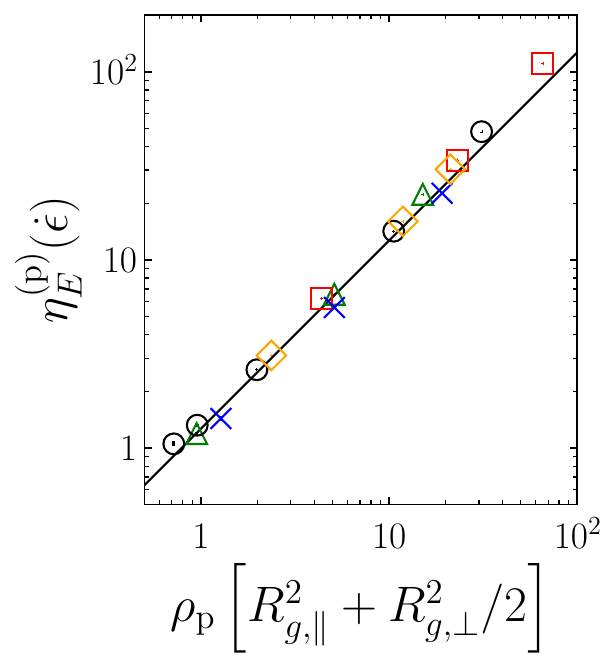} 
  \end{overpic}
  \caption{Polymer contribution $\eta_E^{\mathrm{(p)}}(\dot{\epsilon})$ to the extensional viscosity as a function of $\rho_\mathrm{p} [R_{g,\parallel}^2+R_{g,\perp}^2/2]$ for $(N_\mathrm{p},k_BT,\phi)=(50,1,0.025)$~(green triangle), $(50,1,0.05)$~(black circle), $(50,1,0.1)$~(red square), $(50,0.9,0.05)$~(orange diamond), and $(80,1,0.03)$~(blue cross). The black line indicates $\eta_E^{\mathrm{(p)}}(\dot{\epsilon}) \propto \rho_\mathrm{p} [R_{g,\parallel}^2+R_{g,\perp}^2/2]$. The error bars denote the standard deviations from three independent simulations but are too small to be visible.}
  \label{fig:vis_gyration_polymer}
\end{figure}

In this appendix, we demonstrate that the relation~[Eq.~\eqref{eq:vis_mono_gyration}] of the extensional viscosity for the Rouse-type model~\cite{uneyama2025radius} is applicable to polymer solutions in DPD simulations.
We employ linear polymer chains consisting of $N_\mathrm{p}$ particles connected by FENE springs with the same parameters as those in previous studies~\cite{Jiang2007-rf,Koide2023-ao}. 
We expect that Eq.~\eqref{eq:vis_mono_gyration} holds for monodisperse polymer solutions.
Figure~\ref{fig:vis_gyration_polymer} shows the contribution $\eta_E^{(\mathrm{p})}(\dot{\epsilon})(=\eta_E(\dot{\epsilon})-\eta_E^{(\mathrm{w})})$ of polymers to $\eta_E(\dot{\epsilon})$ as a function of $\rho_\mathrm{p} [R_{g,\parallel}^2+R_{g,\perp}^2/2]$.
Here, $\rho_\mathrm{p}$ denotes the number density of polymer particles.
Data obtained for various $\phi$, $k_BT$, $N_\mathrm{p}$ and $\dot{\epsilon}$ collapse onto a single line $\eta_E^{(\mathrm{p})}(\dot{\epsilon})\propto \rho_\mathrm{p} [R_{g,\parallel}^2+R_{g,\perp}^2/2]$.
Thus, given the assumption that $\zeta$ is constant, we confirm that the relation between the extensional viscosity and the gyration radius~[Eq.~\eqref{eq:vis_mono_gyration}] is applicable to polymer systems in DPD simulations.
%   --------------------
% \begin{suppinfo}

% A listing of the contents of each file supplied as Supporting Information
% should be included. For instructions on what should be included in the
% Supporting Information as well as how to prepare this material for
% publications, refer to the journal's Instructions for Authors.

% The following files are available free of charge.
% \begin{itemize}
%   \item Filename: brief description
%   \item Filename: brief description
% \end{itemize}

% \end{suppinfo}

% \bibliographystyle{achemso}
% \bibliography{my_bib}
\providecommand{\latin}[1]{#1}
\makeatletter
\providecommand{\doi}
  {\begingroup\let\do\@makeother\dospecials
  \catcode`\{=1 \catcode`\}=2 \doi@aux}
\providecommand{\doi@aux}[1]{\endgroup\texttt{#1}}
\makeatother
\providecommand*\mcitethebibliography{\thebibliography}
\csname @ifundefined\endcsname{endmcitethebibliography}
  {\let\endmcitethebibliography\endthebibliography}{}

\end{document}